\documentclass[fleqn,usenatbib]{mnras}

\usepackage{newtxtext,newtxmath}
\renewcommand{\vec}[1]{\ensuremath{\boldsymbol{#1}}}
\newcommand{\nvec}[1]{\ensuremath{\boldsymbol{\hat{#1}}}}
\newcommand{\HL}[1]{\textcolor{black}{#1}}

\usepackage[T1]{fontenc}

\DeclareRobustCommand{\VAN}[3]{#2}
\let\VANthebibliography\thebibliography
\def\thebibliography{\DeclareRobustCommand{\VAN}[3]{##3}\VANthebibliography}

\usepackage{graphicx}	% Including figure files
\usepackage{amsmath}	% Advanced maths commands

\usepackage{siunitx}
\newcommand\T{\rule{0pt}{2.6ex}}       % Top strut
\newcommand\B{\rule[-1.2ex]{0pt}{0pt}} % Bottom strut
\newcommand{\minone}{$^{-1}$}
\newcommand{\tenpow}[1]{$10^{#1}$}
\newcommand{\meyr}{$M_\oplus$~yr\minone}
\newcommand{\mearth}{$M_\oplus$}
\newcommand{\msun}{$M_\odot$}
\newcommand{\ergps}{erg~s$^{-1}$}
\title[Effect of radiation pressure on PPD dispersal]{The effect of radiation pressure on \HL{the dispersal of photoevaporating discs}}

\author[A. Robinson et al.]{
Alfie Robinson,$^{1}$\thanks{E-mail: a.robinson21@imperial.ac.uk}
James E. Owen$^{1}$ and
Richard A. Booth$^{2}$ 
\\
$^{1}$Astrophysics Group, Imperial College London, Prince Consort Road, London SW7 2AZ, UK\\
$^{2}$School of Physics and Astronomy, University of Leeds, Leeds, LS2 9JT, UK
}

\date{Accepted XXX. Received YYY; in original form ZZZ}

\pubyear{2015}

\begin{document}
\label{firstpage}
\pagerange{\pageref{firstpage}--\pageref{lastpage}}
\maketitle

\begin{abstract}
Observed IR excesses indicate that protoplanetary discs evolve slowly for the majority of their lifetime before losing their near- and mid-IR excesses on short timescales. Photoevaporation models can explain this “two-timescale” nature of disc evolution through the removal of inner regions of discs after a few million years. However, they also predict the existence of a population of non-accreting discs with large cavities. Such discs are scarce within the observed population, suggesting the models are incomplete. We explore whether radiation-pressure-driven outflows are able to remove enough dust to fit observations. We simulate these outflows using \textsc{cuDisc}, including dust dynamics, growth/fragmentation, radiative transfer and a parameterisation of internal photoevaporation. We find that, in most cases, dust mass-loss rates are around 5-10 times too small to meet observational constraints. Particles are launched from the disc inner rim, however grains larger than around a micron do not escape in the outflow, meaning mass-loss rates are too low for the initial dust masses at gap-opening. \HL{Only systems that have smooth photoevaporation profiles with gas mass-loss rates $>\sim$\num{5e-9}~\msun~yr\minone and disc dust masses $<\sim$1~\mearth at the time of gap opening can meet observational constraints; in the current models these manifest as EUV winds driven by atypically large high-energy photon fluxes.} We also find that the height of the disc's photosphere is controlled by small grains in the outflow as opposed to shadowing from a hot inner rim; the effect of this can be seen in synthetic scattered light observations.

\end{abstract}


%%%%%%%%%%%%%%%%%%%%%%%%%%%%%%%%%%%%%%%%%%%%%%%%%%

%%%%%%%%%%%%%%%%% BODY OF PAPER %%%%%%%%%%%%%%%%%%

\section{Introduction}

From the infrared (IR) excesses associated with young stars in nearby star forming regions, protoplanetary discs are known to live for around a few million years before dispersing on much shorter timescales, becoming debris discs without a large, optically thick near-IR excess \citep{strom1989,sung2009,wyatt2015,hardy2015}. Although not yet fully understood, the initial primordial phase of disc evolution is currently thought to be governed by two physical processes; viscous evolution driven by turbulence \citep{pringle1981, balbushawley1991}, and magneto-hydrodynamical (MHD) winds that carry mass and angular momentum away from the disc system \citep{blandford1982,pudritz1983,bai2013}. Recent work suggests that MHD winds are the more important process given that non-ideal MHD effects lead to large swathes of the disc being too weakly ionised to support the MRI \citep{gammie1996,perezbecker2011}, and there is currently a large amount of study going into the effect of MHD winds on disc evolution \citep{tabone2022,somigliana2023,weder2023,coleman2024}. 

The secondary, ephemeral phase of disc evolution has generally been understood to be a consequence of photoevaporation of the disc due to stellar irradiation, however MHD winds under certain conditions can also lead to dispersal on fast timescales. Photoevaporation occurs when high energy photons from the central star heat up gas in the upper layers of the disc to energies at which it is able to escape the gravitational potential of the system \citep{hollenbach1994}. A simple “two-timescale” picture of disc evolution and dispersal emerges from comparing the temporal evolution of wind mass-loss rates and the mass accretion rate given by viscous/MHD wind processes, first suggested by \cite{clarke2001}. In this picture, after a few millions of years of primordial evolution, the temporally decaying accretion rate falls below the wind mass-loss rate. This allows the wind to open a gap in the disc at the wind launch radius, as it removes gas on shorter timescales than those required to replace it via the accretion flow from the outer disc. This splits the disc into an inner and outer component. Without the replenishment of material from the outer disc, the inner disc drains rapidly onto the central star in $\sim$ 10$^5$ years. The outer portion of the disc is then removed from the inside-out by the photoevaporative wind. For MHD winds, it has been shown that this two-timescale behaviour can also be found if the magnetisation of the disc increases with time \citep{armitage2013,bai2016}. 

In the photoevaporative case, the loss of the disc's inner regions moves it into the class of discs referred to as `transition' discs; those whose spectral energy distributions (SEDs) lack a hot, near-IR component, indicating an inner hole in the disc material \citep[see e.g.][]{strom1989,owenclarke2012,ercolano2017}. Whilst transition discs are a diverse population best explained by a variety of processes including gap-opening by nascent planets \citep{artymozicz1996,vandermarel2018}, grain growth \citep{dullemond2001}, and dead-zone trapping \citep{regaly2012}, \HL{a small fraction of transition discs are not observed to be accreting and may be best explained through photoevaporation \citep{cieza2010,vdmarel2016}}. Along with direct observations of disc photoevaporation occuring \citep{ercolano2016,weber2020,rab2022}, the loss of the inner disc in the photoevaporation scenario makes it favourable for explaining the observed two-phase nature of disc evolution. 

Photoevaporative flows can be driven by a range of photon energies; extreme ultraviolet (EUV) \citep{hollenbach1994, font2004,alexander2006}, far ultraviolet (FUV) \citep{richling2000,gorti2009}, and X-rays \citep{owen2011,picogna2019,ercolano2023}. Each of these energy regimes exhibit different wind mass-loss profiles amongst other wind properties, but all still produce the same qualitative two-timescale picture described earlier. This is because in all regimes the wind mass-loss rate does not decrease with disc mass as rapidly as the accretion rate; a situation which always leads to gap-opening. 

However, \cite{owen2011} found that models of internal photoevaporation lead to a tension between the simulations, which predict a population of non-accreting discs with large inner cavities and still a substantial mid-IR excess (dubbed ``relic'' discs), and the observations, which exhibit a scarcity of such systems \citep{hardy2015}. Photoevaporation simulations naturally produce the relic disc population due to the fact that the stellar radiation intercepted by the disc decreases with radius whilst the outer disc still retains a large proportion of its initial mass, leading to long timescales associated with the dispersal of the outer disc. The paucity of such systems in the observed population implies the existence of some additional dispersal mechanism currently lacking from the photoevaporation models. 

The solution to the relic disc problem is currently not understood. One proposed solution is an instability termed `thermal sweeping' \citep{owen2012,owen2013}. Thermal sweeping is a process whereby an envelope of vertically expanding gas allows deeper X-ray penetration to the disc mid-plane and, therefore, subsequently rapid ($10^3-10^4$ yr) clearing of the outer disc. However, \cite{haworth2016} reevaluated the criterion for the onset of thermal sweeping and found it to only be effective for discs with large initial accretion rates and low viscosities, suggesting it is not applicable to a large proportion of discs. This process could, however, be revisited to better understand its impact on disc dispersal, particularly by including the effect of FUV heating, not included in the work done by \cite{owen2013,haworth2016}.

Alternatively, \cite{owen2019} suggested that radiation pressure may be able to remove dust from the disc on shorter timescales than the gas evolution. This would resolve the problem because the observational constraints that limit the abundance of relic disc systems are mainly set by observations of the dust continuum emission. If one were to remove the observable dust component of the disc in a shorter timescale, this would lead to congruence with the observed relic disc fraction. The radiation field from the central star exerts an additional force on dust grains in the upper regions of the disc \citep{taklin2003}. This radiation field imparts a purely radial force on any dust grains that experience direct stellar illumination. \cite{owen2019} studied this process in dispersing discs, finding that dust outflows driven by radiation pressure appear to be able to remove the observable dust component within the required timescales. In this work, mass-loss rates of micron-sized grains associated with the radiation pressure outflow were calculated in a 2D (radial and vertical) region localised to the dust-trap at the inner edge of the outer disc (the region in which the turnover in the gas pressure leads to the concentration of dust grains). These mass-loss rates were then included in a 1D disc evolution model that was run for secular ($\sim$~million year) timescales. However, some potential issues with this approach were that the global structure of the disc was not accounted for, and dust growth and fragmentation were neglected. Without including the entire disc extent, especially the vertical structure, it is impossible to know if the outflowing dust actually escapes the disc system or re-enters at larger radii; a question especially important considering we are dealing with a radial force and a disc that may exhibit flaring due to the temperature profile.

In the work presented in this paper, we improve on previous work by moving to a 2D (radial and vertical) framework for both the entire extent of the disc and for the entire simulation time needed to evaluate whether radiation-pressure-driven outflows can remove dust in the timescales required for meeting observational constraints. To do this, we use a newly developed disc evolution code, \textsc{cuDisc} \citep{robinson2024}, which allows the study of axisymmetric discs in both the radial and vertical directions over a significant fraction of the typical disc lifetime. The code evolves the dynamics of differently-sized dust grains using a finite-volume Godunov approach, dust growth and fragmentation by solving the Smoluchowski equation on a fixed grid of grain sizes, and also self-consistently evolves the resultant disc temperature structure from grain-size and wavelength-dependent dust opacities and the dust spatial structure. For this study, we have added radiation pressure and the impact of different photoevaporation models to this framework, allowing us to study a full picture of disc dispersal. 

In section \ref{scenario}, we outline the radiation pressure dispersal scenario, section \ref{methods} describes our numerical methods, sections \ref{mlrates}, \ref{evores} and \ref{graindist} detail the results of our simulations, and in section \ref{obs} we generate synthetic observations.

\section{Scenario}\label{scenario}

\subsection{Radiation-pressure-driven outflows}

\begin{figure}
    \centering
    \includegraphics[width=0.99\columnwidth]{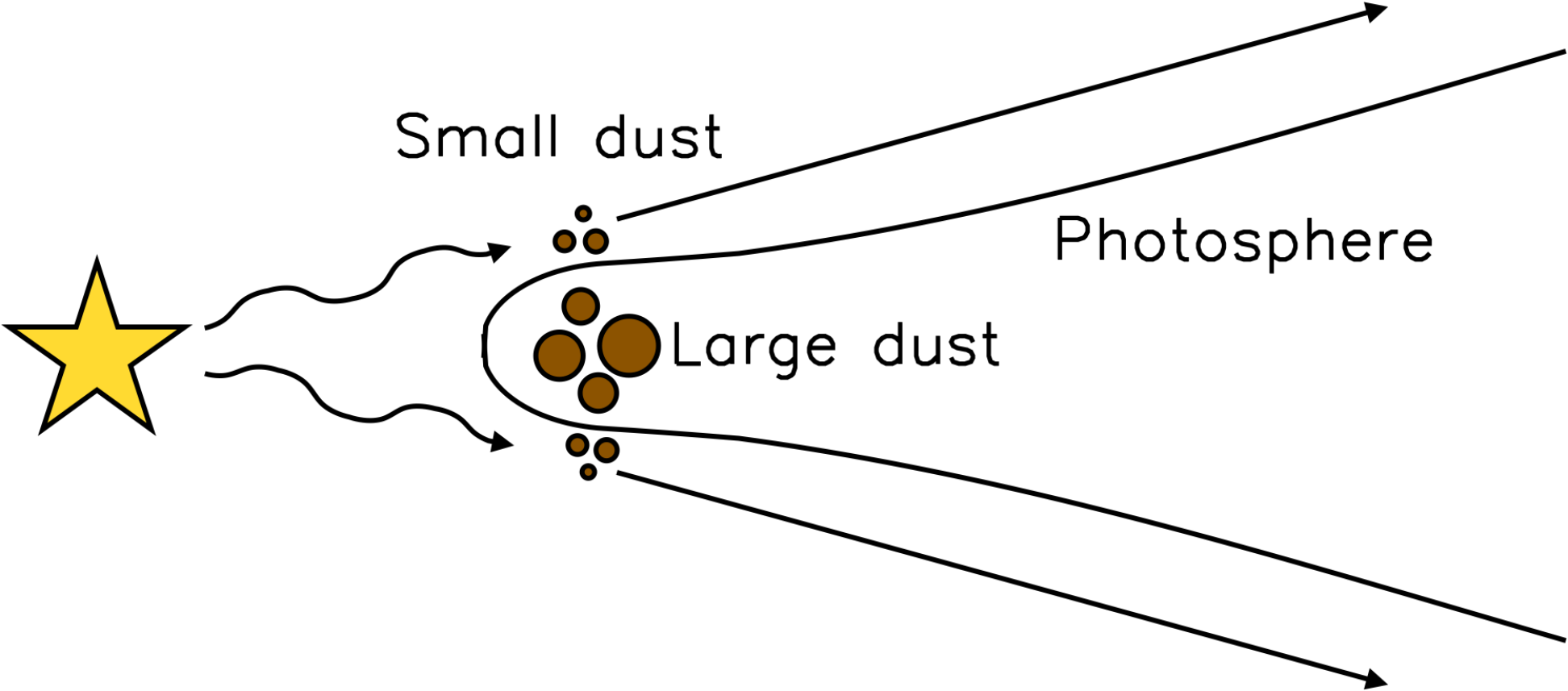}
    \caption{A cartoon depicting the scenario in which dust can be removed from the disc in a radiation-pressure-driven outflow. Small dust grains suspended above the disc photosphere are impinged upon directly by stellar radiation, accelerating them into an outflow. These small grains at the surface must be replenished by fragmentation of large dust grains at the disc mid-plane.}
    \label{radoutflowcartoon}
\end{figure}

Fig.~\ref{radoutflowcartoon} shows a sketch outlining how a radiation-pressure-driven outflow can form. Stellar radiation impinges directly upon grains suspended above the photosphere, exerting a force on them. Small grains inhabit these high altitudes in the disc as they are well-coupled to the gas and therefore held up against gravitational settling by turbulence \citep{dubrulle1995}, and, more importantly in discs with winds, vertical advection due to drag from the wind flow \citep{giacalone2019,hutchison2021,booth2021}. Large grains, in contrast, settle to the disc mid-plane. 

\begin{figure}
    \centering
    \includegraphics[width=0.99\columnwidth]{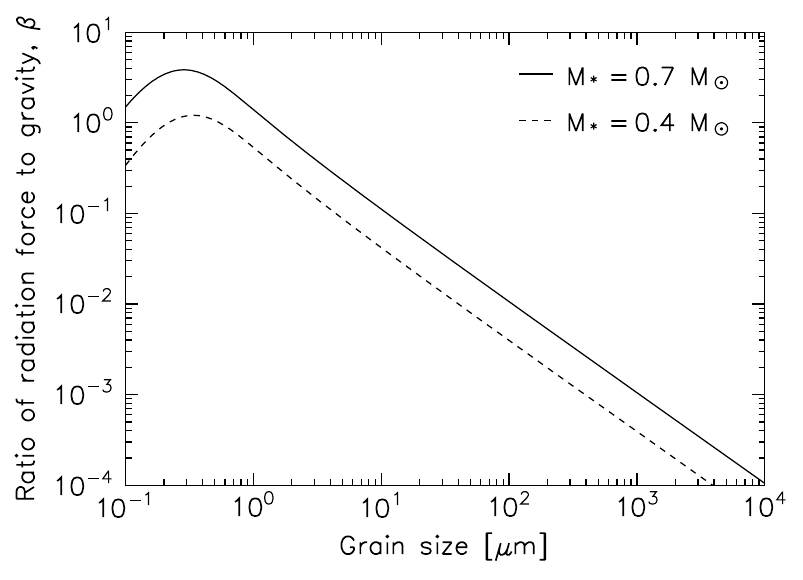}
    \caption{The $\beta$ parameter for different grain sizes and the two stellar types used in our suite of simulations. The effective temperatures and stellar radii that are appropriate for young stars of the shown mass can be seen in Table \ref{tab_params}.}
    \label{betas}
\end{figure}

The ability of radiation pressure to remove dust grains from the dust-trap can be quantified by comparing it to the gravitational force that acts against outward radial forces. The ratio of these forces, the $\beta$ parameter, is given by

\begin{equation}
    \label{radbeta}
    \beta(a) = \frac{1}{4 \pi G M_* c} \int_0^\infty L_*(\nu) \kappa_{\text{ext}}(\nu, a) d\nu, 
\end{equation}
as a function of grain size, $a$, where $L_*$ and $M_*$ are the stellar luminosity and mass, and $\kappa_\text{ext}$ is the wavelength- and grain-size-dependent extinction opacity. \HL{In this work, we assume a blackbody stellar spectrum at all wavelengths apart from in the EUV/X-ray, in which the respective luminosities are varied for our parameter study.} For typical dust compositions \citep[i.e. those used in][]{DSHARP2018}, the opacity peaks at $a\sim0.1-1$~$\mu$m before falling off at a rate of $\sim a^{-1}$ for larger grains. Fig.~\ref{betas} shows how $\beta$ varies with grain size for the two stellar masses we considered in this study (see table \ref{tab_params}), using Planck mean opacities calculated with the DSHARP properties using the DSHARP python package \citep{DSHARP2018,til_birnstiel_2018_1495277}. 

For a system with no gas, a dust grain initially in a Keplerian orbit would need $\beta\geq0.5$ to overcome the gravitational potential and escape the system. However, for systems with gas - and therefore drag - the additional radial force from radiation pressure allows dust grains to travel at orbital velocities that are potentially slower than the gas and remain in a stable orbit, meaning that the grains can be dragged along by the gas and receive angular momentum; for a micron-sized grain at a few au with a $\beta$ of 0.1 this leads to an outward drift velocity of $\sim1000$~\unit{\cm\per\second}. This is a situation essentially opposite to the standard inward drift due to the pressure support felt by the gas \citep{taklin2003}. This outward drift means dust grains can be removed from the system for $\beta\geq0.1$. Given this, we can see from Fig.~\ref{betas} that grains with radii $\leq\sim10$~$\mu$m are affected by radiation pressure and can be driven outwards along radial trajectories.

\begin{table*}
    \centering
    \begin{tabular}{c c}
        \hline Parameter & Value(s) \T \B \\ \hline
        $\alpha$ & \num{2.5e-3} \T \\
        $\mu$ & \num{2.4} \\
        Star: $M_*, T_*, R_*$ & 0.7 $M_\odot$, 4500 K, $1.7$ $R_*$,\hspace{10pt}0.4 $M_\odot$, 3800 K, $1.1$ $R_*$ \\ 
        $v_\text{frag}$ & 5 - 20 \unit{\metre\per\second} \T \B \\
        Grain internal density, $\rho_m$ & 1.68~\unit{\gram\per\cm\cubed} \citep{DSHARP2018}, 1.19~\unit{\gram\per\cm\cubed} \citep{ricci2010}  \\
        Photoevaporation model & \begin{tabular}{c} X-ray O11 (constant column) \citep{owen2011} \\ X-ray P19 (variable column) \citep{picogna2019} \\ EUV \citep{font2004,alexander07} \end{tabular} \\ \hline
    \end{tabular}
    \caption{The various parameters and their values used in our suite of simulations. The stellar parameters ($T_*, R_*$) for a given stellar mass were found using the MIST grid of stellar evolution models \protect\citep{choi2016}. X-ray O11 \& P19 both take X-ray luminosity ($L_X$ in erg~\unit{\per\second}) as their input parameter, whilst EUV model takes the EUV ionising flux ($\Phi$ in photons~\unit{\per\second}) as its parameter.}
    \label{tab_params}
\end{table*}

\subsection{Dispersing the dust disc}\label{dispersalcriteria}

If radiation pressure can drive dust outflows, we then must answer the question of whether or not these outflows can disperse the dust component of the disc on timescales that fit the observations. The currently observed disc SED population implies that outer disc dispersal must occur on timescales that are $\sim~1-10$~\% of the disc lifetime \citep{hardy2015}. As discs typically live for a few Myr, the removal of the majority of the dust must occur within a few hundred thousand years. There are several conditions that have to be met in order for radiation pressure outflows to fulfil such constraints. These are:

\begin{enumerate}
    \item The mass-loss rate of dust from the dust-trap must be large enough to make the disc optically thin to mid-IR within the observational time constraints.
    \item The trajectories of dust grains accelerated at the radius of the dust-trap must point out of the disc system.
    \item The small dust above the disc photosphere must be replenished on timescales fast enough to not inhibit the mass-loss rate due to radiation pressure.
\end{enumerate}

The mass-loss rates required to meet the first condition depend on the mass of dust within the optically thick dust-trap at the inner edge of the disc, as this produces large mid-to-far-IR excesses that are incommensurate with observations of non-accreting transition discs. To remove this observational component, dust must be removed from the trap until it becomes optically thin to mid-IR. From our own simulations and previous work \citep[see e.g.][]{owen2019}, the mass of dust in the trap just after inner disc dispersal is typically $\sim0.1-1~M_\oplus$ for a range of radial drift efficiencies - combining this with observational constraints of $\sim$\tenpow{5}~years gives us required mass-loss rates from the trap of $\sim$\tenpow{-6}\meyr. 

\begin{figure}
    \centering
    \includegraphics[width=0.9\columnwidth, trim={0cm 0cm 0cm 0cm}, clip]{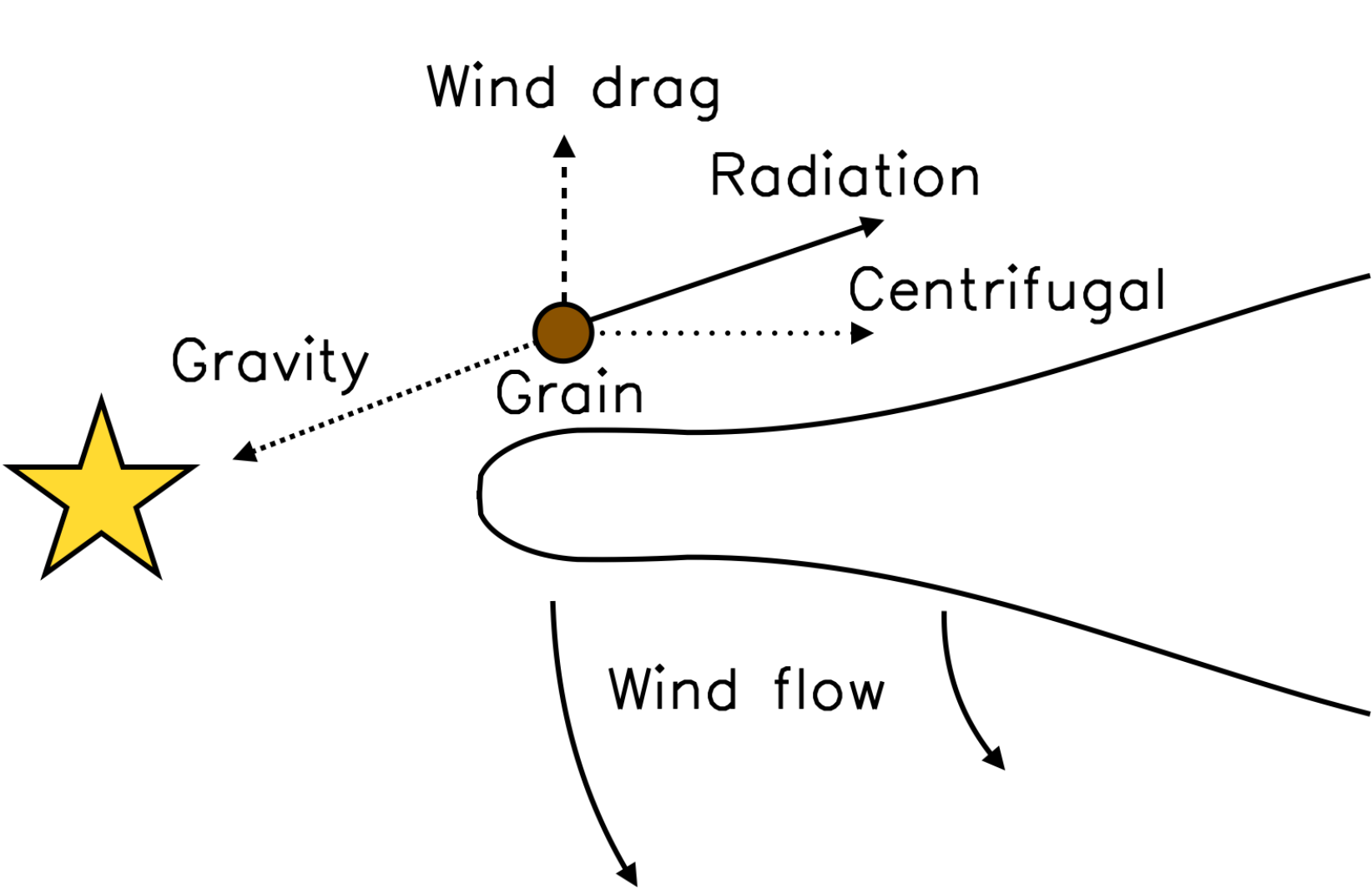}
    \caption{The forces acting on a dust grain suspended above the photosphere of the disc. Due to the cylindrically-radial centrifugal force on the grain, vertical drag from the photoevaporative wind is required in order for the resultant force on the grain to point at an angle above the mid-plane that is greater than the angle made between the mid-plane and the spherically-radial unit vector, $\nvec{r}$. This is important as the disc surface will lie along $\nvec r$ if there is no flaring, and would be even steeper with radius if the disc exhibits flaring.}
    \label{forces}
\end{figure}

Fulfilling the second condition essentially depends on the direction of the resultant force on the dust grains and the disc flaring. Fig.~\ref{forces} shows the forces acting on dust grains above the disc photosphere. Gravity and radiation pressure both act along the spherically-radial line connecting the star and the grain, but in opposite directions. The grain also experiences an outward cylindrically-radial centrifugal force due to its angular velocity. With just these three forces, a grain would have a resultant force that points at an angle above the mid-plane that is less than the angle made between the mid-plane and the spherically-radial unit vector, $\nvec{r}$. For a disc that exhibits no flaring, the disc aspect ratio ($H/R$) is constant, whilst a flared disc has an aspect ratio that increases with radius. This means that for a disc with no flaring, the disc ``surface'' lies along $\nvec{r}$. All of this means that the resultant force on the grain must point at an angle above the mid-plane that is \emph{at least} as large as the angle made between the mid-plane and $\nvec{r}$; otherwise, the grain would ``re-enter'' the disc at some radial distance away from the inner rim, stopping it from escaping the disc. Therefore, we need an additional force to keep the grains above the disc surface. This can be provided by drag from the photoevaporative wind flow, which we will approximate as being a purely vertical force. In the upper layers of the disc, where the gas is increasingly tenuous, the velocity of this flow becomes of order the gas sound speed. For small, well-coupled dust grains, this high-velocity flow exerts a vertical force that can increase the angle of the resultant force above the mid-plane to greater than $\nvec{r}$. If this is sustained until dust becomes unbound from the disc system, then we have succeeded in fulfilling condition (ii). There is also a feature of transition discs that may aid the escape of dust without the need for large vertical forces; shadowing from the hot inner rim, which can lead to the aspect ratio actually decreasing with radius in the inner disc. Whether dust can escape the disc will, therefore, depend heavily on both shadowing and the wind mass-loss profiles. 

The third condition relates to the mechanisms that control the supply of small grains that can be removed by radiation pressure to the surface layers of the disc. These mechanisms are: fragmentation of large grains in the disc interior and delivery of these fragmentation products to the surface via diffusive and advective processes. As small grains are removed from the surface layers of the disc, the rate at which they must be resupplied must be at least large enough to fulfil condition (i). This requires the fragmentation rate of the largest grains at the mid-plane to be sufficiently high to feed the mass-loss rate of small grains from the surface. By writing the relative velocity due to turbulence as $\sqrt{\alpha \text{St}} c_s$ \citep[see ][]{birnstiel2009}, we can approximate the full form of the fragmentation calculation (presented in Section~5 of \citealt{robinson2024}) to find a simple form for the density production rate of small grains from the largest grains: 

\begin{equation}
    \label{sigma_frag_app}
    \dot{\rho}_\text{frag} = 2m_l n^2_l \sigma_l \sqrt{\alpha \text{St}_l} c_s F_s,
\end{equation}
where $n_l$, $\sigma_l$, $\text{St}_l$ and $m_l$ are the mid-plane number density, collision cross-section, Stokes number and mass of the largest-sized grains respectively, $\alpha$ is the dimensionless turbulence parameter \citep[see][]{shakira1973}, $c_s$ the isothermal sound speed, and $F_s$ the fraction of fragments that are small grains that can enter the outflow. Assuming that 0.1~\mearth of dust must be lost from the trap in a timescale of 100,000 years to make the disc optically thin, and using typical values used in this study, we find that the mid-plane mass volume density ($\rho=mn$) of the largest grains must be greater than $\sim$\tenpow{-15}~\unit{\gram\per\cm\cubed} for collision rates to be large enough to fuel the surface mass-loss. The small grains now at the mid-plane also need to move up to the surface layers at a sufficient rate. Using the definition of the viscous timescale, $t_\nu = Z^2/\nu$, we can approximate the rate of surface density transport via diffusion to be $\rho_{<\mu\text{m}}\alpha H\Omega_K$, where $\rho_{<\mu\text{m}}$ is the mid-plane density of sub-micron dust grains and $\Omega_K$ the Keplerian angular velocity. Using the typical parameters used in this study, we find that the mid-plane mass density of sub-micron-sized grains must be greater than $\sim$\tenpow{-16}~\unit{\gram\per\cm\cubed} in the dust-trap in order for diffusion to deliver enough dust to the surface. In addition to diffusion, advection via drag with the wind also contributes to the delivery of small dust grains to the surface \citep{booth2021}, meaning that delivery of dust grains to the surface will not be a limiting factor. 

As mass is lost from the disc, these criteria will cease to be fulfilled (and actually may need to cease being fulfilled, as a dust density of \tenpow{-15}~\unit{\gram\per\cm\cubed} would give a surface density in a dust-trap at 10 au of a few $\times~$\tenpow{-3}~\unit{\gram\per\cm\squared}, which would likely be optically thick to mid-IR); however, we can use them at the outset of the formation of radiation-pressure-driven outflows to evaluate the expected efficacy of the outflows in dispersing the dust disc within the observational time constraints. We will refer back to these criteria when analysing our results. 

\section{Methods}\label{methods}

To get a full picture of disc evolution under the influence of radiation pressure and various photoevaporation models, we evolve the disc from the start of the primordial (class II) phase up until full dispersal. To achieve this, we split up the evolution into two phases that are modelled in different manners. First, we run simpler 1D models of disc evolution for the primordial phase until the inner disc has drained before switching to a 2D framework for the final period of outer disc dispersal. We run 1D models for the initial phase as these have been shown to do sufficiently well for standard discs \citep{birnstiel2012,robinson2024}, and we, therefore, save on computational resources allowing us to complete a full parameter scan.

\subsection{1D models}

For the 1D models, we use the standard approach used by many authors. We evolve the gas by solving the viscous evolution equation \citep{pringle1981} with an additional source term due to photoevaporation,

\begin{equation}
    \label{gas1D}
    \frac{\partial \Sigma_g}{\partial t} = \frac{3}{R} \frac{\partial}{\partial R}\left[R^{\frac{1}{2}} \frac{\partial}{\partial R}\left(R^{\frac{1}{2}} \nu \Sigma_g \right)\right] - \dot{\Sigma}_\text{wind},
\end{equation}
where $\Sigma_g$ is the gas surface density, $\nu$ the turbulent viscosity, and $\dot{\Sigma}_\text{wind}$ the surface density loss due to photoevaporation. The equation is solved in two stages via operator splitting; in the first stage, the first term on the right-hand side is solved as a diffusion problem using an explicit forward-time-centered-spatial scheme, and in the second stage, the source term update (i.e. wind mass-loss) is solved through explicit Eulerian updates. The wind surface density loss rates are taken from three different works to allow us to compare the effect of different photoevaporation models. We implement both X-ray and EUV models. Given that the 1D simulations run until gap-opening, we use the primordial (no inner cavity) disc wind profiles. These models and their respective references can be found in Table \ref{tab_params}. 

For the dust, we use the two-population model detailed by \cite{birnstiel2012} and solve a single advection-diffusion equation for the dust,

\begin{equation}
    \label{dust1D}
    \frac{\partial \Sigma_d}{\partial t} + \frac{1}{R} \frac{\partial}{\partial R} \left\{R \left[ \Sigma_d \Bar{u} - \frac{\nu \Sigma_g}{\text{Sc}} \frac{\partial}{\partial R} \left(\frac{\Sigma_d}{\Sigma_g} \right) \right] \right\} = 0,
\end{equation}
where $\Sigma_d$ is the total dust surface density and $\Bar{u}$ is the mass-weighted dust advection velocity. This velocity is an average over two dust populations; one representing the ``small'' grains, and the other the ``large'' grains. The grain radius representing the large grains is a function of radius, set as the minimum of the maximum grain sizes allowed by either radial drift or fragmentation due to turbulence or relative drift velocities, whilst the small grain radius is set as the monomer size, chosen by us to be 0.1~\unit{\micron}. Sc is the Schmidt number, a dimensionless number that represents the strength of dust-gas coupling. For more specific details of the model, we refer the reader to \cite{birnstiel2012}. We solve the dust equation using a second-order finite-volume Godunov scheme in the manner of \cite{STONE2009139}, with the second-order primitive reconstruction performed using a van Leer flux limiter \citep{VANLEER1974361}. More detail on this method can be found in \cite{robinson2024}, as we solve the 1D dust equation using the same method used by the 2D algorithm for \textsc{cuDisc}. To allow the dust to flow through the inner boundary without unphysical build-ups, we use outflow boundaries for the gas and set the boundary cell densities to those in the first active cell (to avoid steep positive pressure gradients that lead to dust-trapping) and set the dust density in the boundary cells to the floor value (\num{1e-15}~\unit{\gram\per\cm\squared}). We neglect dust loss due to wind entrainment in these simulations as other works have argued that this is small during the primordial phase \citep{owen2011a}.

The turbulent viscosity, $\nu$, is chosen to be consistent with the photoevaporation studies that we are using for our wind profiles. In these, the viscosity is linear with respect to radius, i.e. $\nu~=~\nu_0(R/\text{au})$. This was chosen as the nature of the underlying cause of viscosity is not well constrained whilst a linear relationship with radius is fairly consistent with observations \citep[see e.g.][]{hartmann1998}. In our simulations, $\nu_0$ is calculated using the $\alpha$ formalism \citep{shakira1973}, $\nu_0=\alpha c_{s,au}^2/\Omega_{K,au}$~\unit{\cm\squared\per\second}, where $c_{s,au}$ is the sound speed at 1 au assuming a mid-plane temperature of 100 K at 1 au, and $\Omega_{K,au}$ is the Keplerian angular velocity at 1 au. We use this viscosity for both the 1D and 2D simulations.

\begin{figure}
    \centering
    \includegraphics[width=0.95\columnwidth]{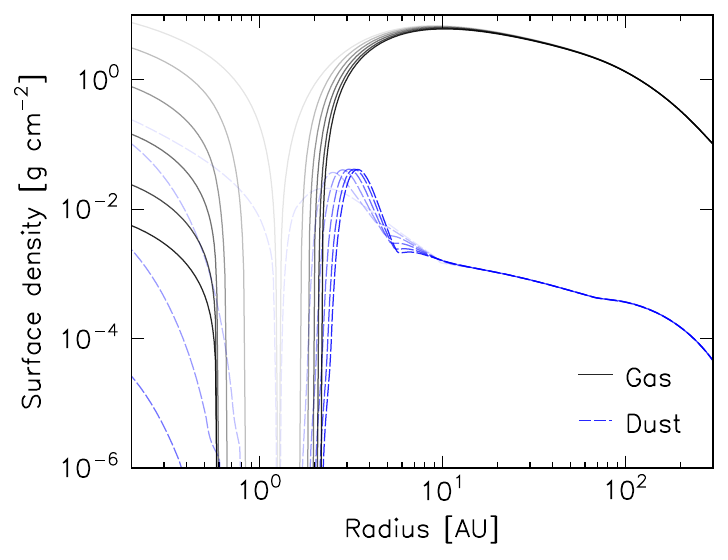}
    \caption{A 1D simulation of disc evolution from \HL{the time of }gap-opening until inner dust disc drainage. For this run, the gap opened after 3.6~Myr, and the inner dust disc had fully drained after 48,000~years. The opacity of the lines increases with time, with snapshots at 0, 10, 20, 30, 40 and 48~kyr. The \protect\cite{owen2011} X-ray photoevaporation model was used for this run, with an X-ray luminosity of $\log_{10}L_X~=~30$. }
    \label{1drun}
\end{figure}

We set up each 1D simulation with 500 logarithmically-spaced radial cells between 0.1 and 500~au. We initialise the gas surface density in each 1D simulation with the zero-time self-similar solution derived by \cite{lbp74} with a characteristic radius of 30 au and an initial gas mass of 0.07 $M_\odot$. A range of initial conditions for each of the wind mass-loss mechanisms have been explored in previous works \citep[see e.g.][]{alexander2006,owen2011,picogna2019}, but we use these initial conditions for all simulations as our final conclusions are not sensitive to them, and therefore varying them is not important for our exploration. We initialise the dust surface density as being well-mixed with the gas with a dust-to-gas ratio of 0.01, i.e. $\Sigma_d(R)=0.01\Sigma_g(R)$. An example of a 1D simulation can be seen in Fig.~\ref{1drun}, which is essentially a repeat of previous simulations in \cite{owen2019}. After \HL{the gap opens}, the inner dust disc drains in $\sim50,000$~years in this particular simulation, and we can see the expected approximately Gaussian profile of the build-up of dust in the pressure trap at the outer disc's inner rim. 

\subsection{2D models}

After completion of the 1D models, we use \textsc{cuDisc} \citep{robinson2024} to continue the simulations. \textsc{cuDisc} is an axisymmetric disc evolution code that includes dust dynamics, growth and fragmentation, and radiative transfer in both the radial and vertical dimensions. \textsc{cuDisc} employs a second-order finite-volume Godonuv solver for dust dynamics, solves the Smoluchowski equation for dust growth and calculates radiative transfer using a multi-frequency hybrid ray-tracing/flux-limited-diffusion method; see \cite{robinson2024} for full details. The simulation grid uses cells that are bounded by lines of constant cylindrical radius, $R$, and lines of constant angle above the mid-plane, $\theta$; however, we will generally use cylindrical $Z$ when discussing or plotting the vertical structure of the simulations. \textsc{cuDisc} does not evolve the gas in 2D as full hydrodynamics would be too computationally expensive to allow simulations to be run for Myr timescales, and therefore the gas is still evolved in 1D using the method described in the previous section. The 2D gas density is calculated through hydrostatic equilibrium. We implement the effects of winds and radiation pressure, as detailed in the following sections.

\subsubsection{Photoevaporative winds}\label{PEwindsimp}

As the 2D models commence once the inner disc has drained, we use the wind mass-loss profiles for discs with inner cavities given by the works listed in Table \ref{tab_params}. These profiles depend on the radius of the inner cavity. To calculate the vertical gas velocity, we assume that the surface density loss is supplied by the vertical flow of gas through the disc \citep{booth2021}, and calculate the velocity through mass conservation,

\begin{equation}
    \label{wind_v_Z}
    v_{g,Z} = \frac{\dot{\Sigma}_\text{wind}}{\rho_g}. 
\end{equation}
\HL{There are two caveats to this approach. The first is that we overestimate the velocity in the upper layers of the disc because our calculation of the gas density assumes hydrostatic equilibrium, an assumption that breaks down in the wind region. However, this does not significantly affect the drag force on dust grains, which is dependent on the mass flux of gas, $\rho_g v$, and therefore does not affect whether dust is entrained. The second is that at the inner rim of the disc, the wind is launched radially inwards before redirecting vertically upward \citep{alexander2006a,owen2010}, which is not included in our assumption of a purely vertical wind; the effect of which is discussed in section \ref{mlrates}.}

\subsubsection{Radiation pressure}\label{radpressmethod}

Radiation pressure is included as an additional source term to the momentum equations and calculated in the explicit update alongside gravitational and curvature terms. For dust species $k$ in a given cell, the force exerted per unit volume is given by

\begin{equation}
    \label{radpress}
    \vec f_{\text{rad},k}= \int_0^\infty \frac{L_{*,\nu}}{4\pi c r^2} \text{e}^{-\tau_{*,\nu}} \rho_{d,k} \kappa_{\text{ext},k,\nu} d\nu \, \vec{\Hat{r}}, 
\end{equation} 
where $r$ is the spherically-radial distance from the star to the cell, $c$ the speed of light, $\rho_{d,k}$ the mass density of dust species $k$, $L_{*,\nu}$ the frequency-specific stellar luminosity, $\tau_{*,\nu}$ the frequency-specific total optical depth between the star and the cell and $\kappa_{\text{ext},k,\nu}$ the frequency-specific extinction in the cell. We cover our choice of opacities in Section \ref{opacities}. In practice, this integral is carried out as a sum over the wavelength bins used in the simulation,

\begin{equation}
    \label{radpress_discrete}
    \vec f_{\text{rad},k}= \sum_B \frac{L_{*,B}}{4 \pi c r^2} \text{e}^{-\tau_{*,B}} \rho_{d,k} \kappa_{\text{ext},k,B} \vec{\Hat{r}}, 
\end{equation} 
where $B$ indexes the wavelength bands. The total optical depth to the star in cell $ij$, $\tau_{*,B,ij}$, is calculated via 

\begin{equation}
    \label{tau_to_star}
    \tau_{*,B,ij} = \sum_{l=0}^{l=i} \sum_k \rho_{d,ljk}\kappa_{\text{ext},ljk,B} + \rho_{g,lj}\kappa_{g_\text{ext},lj,B},
\end{equation}
where $\rho_{g,lj}$ and $\kappa_{g_\text{ext},lj,B}$ are the gas mass density and extinction (in wavelength band $B$) in cell $lj$ respectively.

\subsubsection{Opacities}\label{opacities}

To calculate grain size and wavelength-specific opacities, we use the Python package for the DSHARP Mie-Opacity Library \citep{til_birnstiel_2018_1495277}, detailed in \cite{DSHARP2018}. With this, we calculate both the absorption and scattering opacities for a wavelength range of $0.1-10^5$~\unit{\micron} and a grain size range of $0.1$~\unit{\micron}$-50$~cm. We primarily use the DSHARP grain composition which includes a 20\% mass fraction of water ice, giving an internal grain density of 1.68~\unit{\gram\per\cm\cubed}. For comparison, we also run some simulations using the composition detailed by \cite{ricci2010}, which includes a 35\% mass fraction of water ice, giving an internal grain density of 1.19~\unit{\gram\per\cm\cubed}. The optical constants used for the Mie calculations are taken from \cite{henning1996,zubko1996,draine2003,warren2008}.

\subsubsection{Grid parameters and initial conditions}\label{setupICS}

For each 2D simulation, we use a spatial grid with 200 logarithmically-spaced radial cells and 200 vertical cells, spaced according to a power-law, $\theta^n$, with $n=0.75$. The power-law spacing concentrates vertical cells about the mid-plane, allowing us to resolve the well-settled large dust grains. We use reflecting boundary conditions at the mid-plane, meaning that we can set our inner vertical boundary to $\theta_\text{min}=0$ ($\theta=0$ is defined as the mid-plane in \textsc{cuDisc}), and use an outflow boundary condition at the outer vertical boundary, setting $\theta_\text{max}=\pi/6$~rad. Both radial boundaries are set with outflow boundary conditions, with the outer boundary set to 300~au for every simulation, as by the time of gap-opening in the 1D simulations, the gas surface density at this radius is at least 4 e-folds lower than the peak value. The radius of the inner boundary is simulation-dependent and is set to be the radius of the inner cavity of the gas disc from the completed 1D simulations, minus 0.25~au. 

The stellar heating and radiation pressure calculations use 200 logarithmically-spaced wavelength bins, spanning a wavelength range of $0.1-10^5$~\unit{\micron}. These wavelengths, and the opacities associated with them, are binned into 20 logarithmically-spaced wavelength bands for the flux-limited-diffusion (FLD) calculations used for radiative transport of re-emitted thermal radiation. A constant background temperature of 10~K is applied to provide a baseline radiation field that bathes the entire disc. We set the relative tolerance of the FLD solver to $10^{-5}$. 

For the dust, we use a size grid with 150 logarithmically-spaced bins between 0.1~$\mu$m and 50~cm, giving us a grid resolution of $m_{n+1}/m_{n}=1.36$. This choice ensures that there are at least 7 bins per mass decade, a requirement for accuracy in the coagulation routine \citep{ohtsuki1990}. We set the relative and absolute tolerances for the time-integration of the coagulation solver to $10^{-3}$ and $10^{-10}$, respectively. For the fragment distribution power-law exponent ($\eta$ in Eqn. 66 in \citealt{robinson2024}), we use the standard value of 11/6 \citep{dohnanyi1969,TANAKA1996450} and we set the factor that controls the amount of material that an impactor can remove relative to its mass ($\chi_\text{im}$ in Eqn. 68 in \citealt{robinson2024}) as unity. For dust diffusion, we set the Schmidt number to unity. To avoid numerical issues due to division by small numbers, we use a dust-to-gas ratio floor value of $10^{-12}$, and a gas density floor of $10^{-40}$~\unit{\gram\per\cm\cubed}. We use a dust-to-gas ratio floor as opposed to a dust density floor as it avoids spurious dust diffusion in regions where the dust is floored. \HL{For our fiducial models, we set the dust fragmentation velocity to 10~\unit{\metre\per\second} and use the DSHARP opacities with an associated internal grain density of 1.68~\unit{\gram\per\cm\cubed}}.

We only study the dust dynamics \HL{up to the the wind base. This is sufficient because grains that are delivered to the wind base are generally removed \HL{from} the system \citep{hutchison2021,booth2021}}. We define the wind base, or wind launch surface, as the height in the disc where the spherically-radial column density of neutral gas to the star is equal to the maximum neutral column that ionising radiation can penetrate. We set this to the value for X-ray photons, $10^{21}$~\unit{\per\cm\squared} \citep{ercolano2009}. To avoid short time-steps due to fast-moving dust in regions that do not impact our results, we floor any dust that crosses the wind base. Although the maximum penetrated column can be lower for EUV winds \citep{hollenbach1994}, this choice does not affect our results \HL{because the wind base is far above the photosphere in the EUV simulations due to lower initial dust masses at the time of gap-opening. Dust grains on trajectories that cross the wind base are in a region of very low gas density and experience the maximum possible radiation pressure force meaning they are certain to leave the system.  Thus, the choice of penetration depth does not affect our results.}

\begin{figure}
    \centering
    \includegraphics[width=0.99\columnwidth]{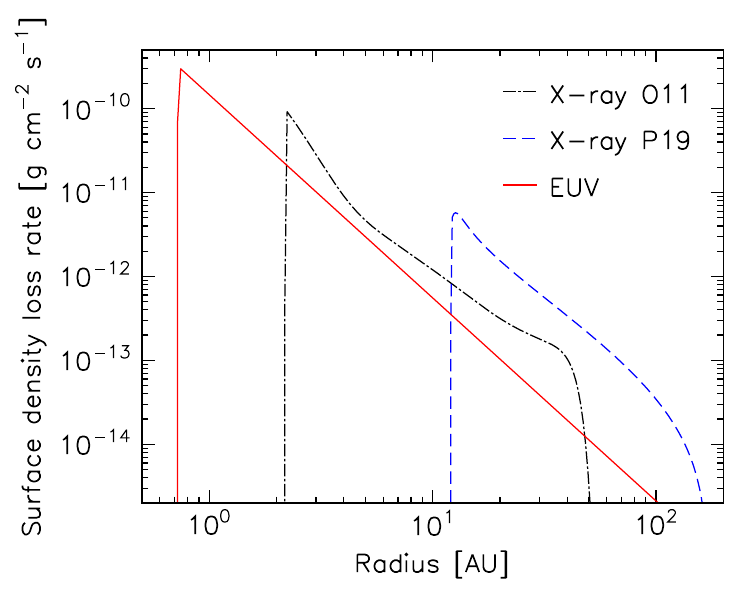}
    \caption{Surface density loss profiles for the different photoevaporative models listed in Table \ref{tab_params}. These profiles are those used in the fiducial 2D simulations detailed in section \ref{mlrates}, and are for direct illumination of the outer disc once a gap has opened in the gas and the inner disc has drained onto the star. The ionising radiation strength is set to the fiducial value for each wind model; the X-ray luminosity, $L_X=\num{1e30}$~\ergps, for the X-ray O11 and B models, and the EUV ionising flux, $\Phi=\num{1e43}$~photons~s$^{-1}$, for the EUV model. The integrated mass-loss rates for the three models are \num{4.9e-9}, \num{1.9e-8}, \num{5.5e-9}~\msun~yr$^{-1}$ for the X-ray O11, P19 and EUV models respectively.} 
    \label{windprofiles}
\end{figure}

\HL{We examine the effects of three different photoevaporation models; the X-ray models of \cite{owen2011} and \cite{picogna2019} (hereafter X-ray O11 and P19 respectively), whose strength is controlled by the X-ray luminosity, $L_X$, and the EUV model presented by \cite{font2004,alexander07}, the strength of which is controlled via the EUV ionising photon flux, $\Phi$. For our fiducial models, we use a 0.7~\msun star (see table \ref{tab_params} for the other associated stellar parameters) and choose values of $L_X=\num{1e30}$~\ergps, $\Phi=\num{1e43}$~photons~s$^{-1}$. Whilst this EUV rate is larger than typical EUV fluxes observed in T Tauri stars \citep{pascucci2014} and those used in previous works \citep[see e.g.][]{alexander2006,alexander07}, it is chosen to give similar gas mass-loss rates to the X-ray models. We do this because, in this study, we aim to understand the conditions necessary for rapid dust removal in photoevaporating discs and therefore want to compare the effect of different wind profiles whilst maintaining realistic disc lifetimes. We also note that recent work by \cite{sellek2024} concludes that the O11 and P19 models overestimate gas mass-loss rates by up to around an order of magnitude; we comment on this when discussing our results. We note that the models by \cite{sellek2024} do not currently include discs with inner holes as are studied in this work. Fig. \ref{windprofiles} shows the surface density loss rates for each wind model in the fiducial simulations. The integrated mass-loss rates for the three models are \num{4.9e-9}, \num{1.9e-8}, \num{5.5e-9}~\msun~yr$^{-1}$ for the X-ray O11, P19 and EUV models respectively. One differentiating feature between the models is how peaked the mass-loss is at the disc inner rim; the EUV model has a smoother profile than both X-ray models, being significantly less peaked than the O11 model.} 

\HL{We choose a fiducial turbulence of $\alpha=\num{2.5e-3}$ to directly compare to previous works that reproduce observed disc lifetimes \citep[see][]{owen2019}. This value is larger than recent observational work implies (\tenpow{-4}$-$\tenpow{-3}, see \citealt{manara2023} for a review), however using a lower value would not affect our qualitative results; this is discussed in section \ref{dmfluxes}.}

To convert from the 1D simulation results to the initial set-up for the 2D simulations, the following steps are taken:

\begin{enumerate}
    \item The multi-species 2D dust density distribution is initialised from the 1D dust surface density by calculating the volume density for dust species $k$ from
    \begin{equation}
        \label{2ddustdensity}
        \rho_{d,k}(R,Z) = {f_k}\frac{\Sigma_d(R)}{\sqrt{2\pi} h_k(R)} \exp\left(-\frac{Z^2}{2h_k^2(R)}\right), 
    \end{equation}
    where $f_k$ is the fraction of the total dust surface density distributed into size-bin $k$. The distribution is set according to an MRN distribution \citep{MRN77} with a sharp upper exponential cut-off at a grain size set as the minimum of the fragmentation and drift limits at a radius $R$ in the disc (see \citealt{birnstiel2012} for the forms of these limits). $h_k$ is the scale-height of dust species $k$, given by
    \begin{equation}
        \label{dustscaleheight}
        h = H \sqrt{\frac{1}{1+\text{St}/\alpha}},
    \end{equation}
    where $H$ is the gas scale height. The dust velocities are initialised in Keplerian orbits with meridional velocities set according to steady-state drift and settling velocities,
    \begin{equation}
        \label{v_inits}
        \begin{pmatrix}
            v_{d,R} \\ v_{d,\phi} \\ v_{d,Z}  
        \end{pmatrix}
        =
        \begin{pmatrix}
            \dfrac{v_{g,R} - \eta \text{St} \Omega_K R}{1+\text{St}^2} \\
            \Omega_K R \\ - \text{St} \Omega_K Z 
        \end{pmatrix},
    \end{equation}
    where $\eta$ accounts for the deviation of the azimuthal gas velocity from Keplerian due to pressure.
    \item The initial disc temperature is calculated through a nested iteration process; we calculate the temperature and update the gas vertical distribution through hydrostatic equilibrium, iterating this procedure until the root-mean-square error in temperature over the entire disc drops below 0.01\%. We then recalculate the dust distribution according to step (i) and iterate over this whole procedure until the dust, gas and temperature distributions are all converged. 
\end{enumerate}
\section{Results: dust mass-loss rates}\label{mlrates}

The combined processes of radiation-pressure-driven dust outflows and global disc evolution create a complicated picture that hasn't been studied before. We begin by looking solely at the features of the dust outflows formed in the inner regions of the disc, and how the mass-loss rates of dust from the dust-trap depend on different disc parameters. We use our findings from this study to understand the evolutionary calculations described in section \ref{evores}.

\HL{To explore how mass-loss rates depend on different disc parameters, we take our fiducial models described in section \ref{setupICS} and modulate each parameter. Our fiducial dust surface densities as those reached at the end of 1D simulations using the fiducial values: these have peak dust-trap surface densities of 0.041, 0.012 and 0.033~\unit{\gram\per\cm\cubed} for the X-ray O11, P19 and EUV models respectively. The respective radii of the gas disc inner rims of each of the fiducial models at the end of the 1D simulations are $\sim2.2$, 12 and 0.7~au. The end points of the fiducial 1D simulations are used as inputs for each 2D simulation. For each parameter, we run 2D simulations for each fiducial value and for 0.5$~\times$, 2~$\times$ and 5~$\times$~the fiducial values. To isolate the dust evolution, we turn off the 1D gas surface density evolution - keeping, however, the vertical gas velocities due to the wind. These 2D models are run for 1000~years without photoevaporative winds or radiation pressure to relax the system, then run with all forces until a quasi-steady state is reached.}

\subsection{Outflow characteristics}\label{outflowcharacteristics}

\begin{figure*}
    \centering
    \includegraphics[width=\textwidth]{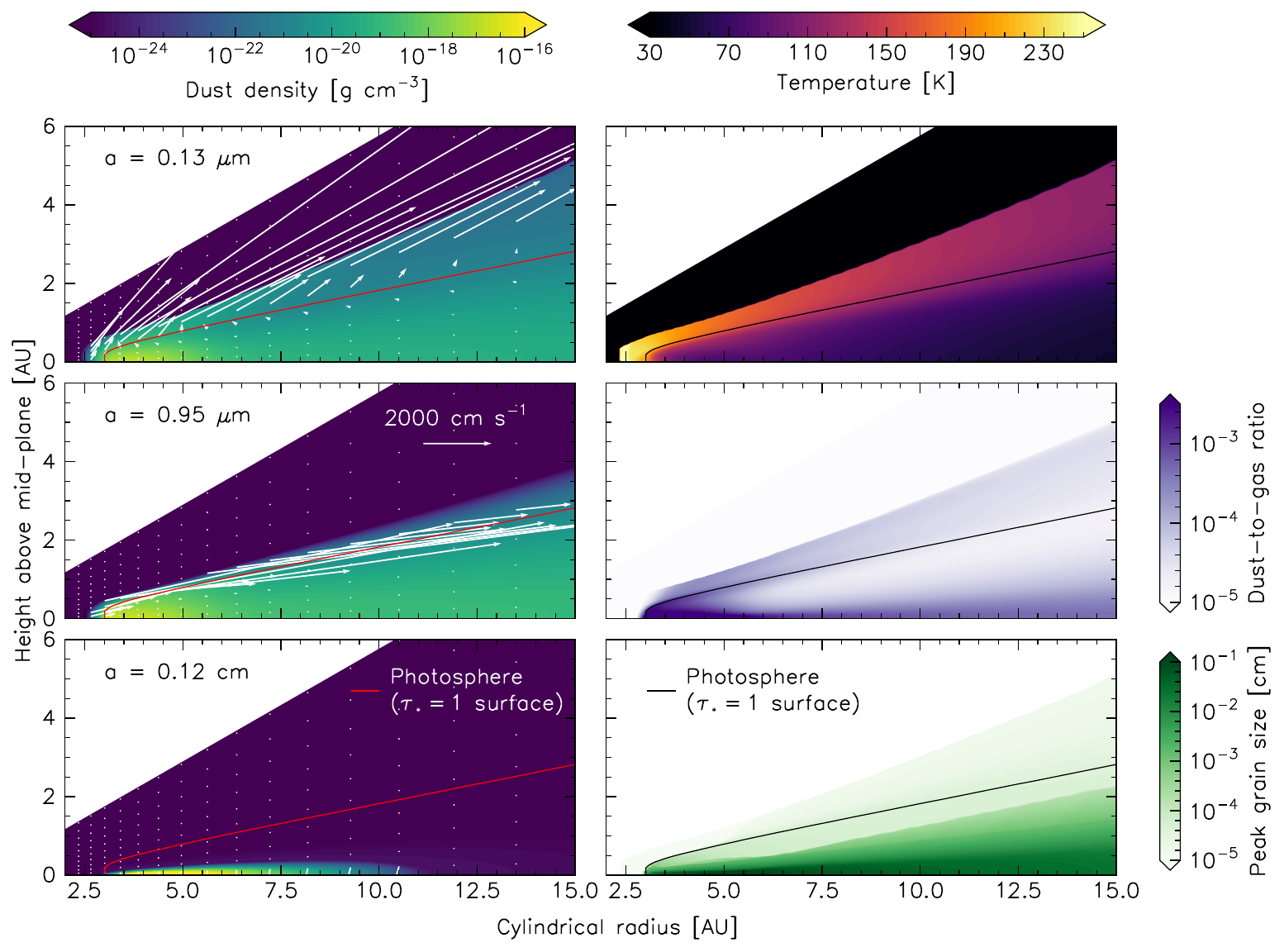}
    \caption{A snapshot of the fiducial X-ray O11 model after 10,000 years. The 2D densities and velocities of 3 different grain sizes are shown in the first column. To find the photosphere, the optical depth is calculated at a wavelength of 0.64~\unit{\micron}, the location of the peak of the stellar blackbody spectrum. From the top down, the second column shows the temperature profile, the dust-to-gas ratio and the grain size associated with the peak of the dust density distribution at each location in the disc. \HL{The temperature profile is shown up to the wind base, and in this figure is set to 0 (black) above it to aid the reader. The region above the wind base is also where the dust density is at the floor value.}}
    \label{2drunexample}
\end{figure*}

Fig. \ref{2drunexample} shows a snapshot of the fiducial X-ray O11 simulation after 10,000 years. From the density profiles of differently-sized grains, we can see the morphology of the outflow; sub-micron-sized grains suspended above the photosphere have high velocities and move along trajectories that point out of the surface of the disc, micron-sized grains above the photosphere are accelerated, but with trajectories that mostly point back below the photosphere, and mm-sized grains are confined to the mid-plane due to settling. The temperature profile exhibits a super-heated disc surface due to the opacity of the small grains that inhabit this region, with a cool mid-plane governed by the opacity of the mid-plane dust to the re-processed stellar radiation emitted by the surface layers. The total dust-to-gas ratio shows that the outflow raises the dust content in the upper layers of the disc to higher concentrations than at intermediate heights, especially at radii slightly exterior to the dust-trap. We also show the grain size associated with the peak of the dust density distribution at each location in the disc. The most noticeable feature of this is the sharp transition from $\sim10$~\unit{\micron} grains to $<1$~\unit{\micron} grains at an altitude slightly below the photosphere; this is because the dust distribution is double-peaked at these heights, one peak at the size of the grains in the outflow, and one at the maximum grain size set by the balance of settling, diffusion, growth and fragmentation within the disc interior. At a certain altitude, the grains in the outflow become the dominant grain size, causing the sharp transition we see in the peak grain-size distribution. This feature of the dust grain-size distribution is important for the location of the disc photosphere; this is discussed in section \ref{graindist} where fig. \ref{graindist_ratesbehindtrap} shows an example grain-size distribution as a function of height.

\begin{figure}
    \centering
    \includegraphics[width=0.99\columnwidth]{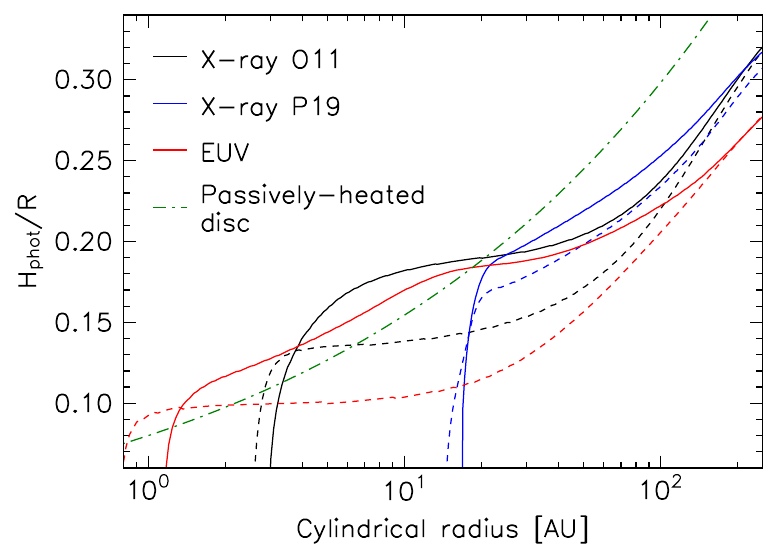}
    \caption{Aspect ratios of the photospheres, $H_\text{phot}/R$, for each fiducial wind model simulation. Dashed lines show the aspect ratios at the start of the simulations, before the outflow has formed, whilst the solid lines are after 4000~years of evolution.}
    \label{hphots}
\end{figure}

Fig. \ref{hphots} shows the photosphere aspect ratios, $H_\text{phot}/R$, for the fiducial simulations of each wind model, both at the start of the simulation, when no dust has been drawn up into the wind or affected by radiation pressure, and after 4000 years of evolution. The photosphere is defined as the surface in the disc at which the optical depth to the star at the peak of the stellar blackbody spectrum ($\sim0.6$~\unit{\micron}) is unity. The dependence of the aspect ratio on radius gives an indication of the amount of disc flaring. For reference, the level of flaring for a passively-heated primordial disc is also shown ($H_\text{phot}/R\propto R^{2/7}$, \citealt{chiang1997}). Initially, all the discs have lower altitude photospheres than at later times, and the discs with inner radii $<10$~au exhibit no flaring in the inner regions of the disc. All models have the same level of flaring as the passively heated disc for radii greater than around 50 au. After the outflow has formed, the X-ray O11 model has the least amount of flaring, up to around 30 au, compared to the other models; the X-ray P19 model exhibits passively-heated levels of flaring at all radii, whilst the EUV model shows an initial level of flaring commensurate with passive-heating before flattening between around 10 and 30 au. 

The difference between the initial photospheres and those where the outflow has formed indicates that the level of flaring in the evolving disc is set by the morphology of the outflow, as opposed to shadowing due to a puffed-up inner rim. Any effects due to shadowing are removed by the ability of the wind and radiation pressure to carry dust up to higher altitudes of the disc. The evolving X-ray O11 model shows less flaring than the X-ray P19 model because the trajectories are much closer to radial due to reduced drag from the wind, which has a lower mass-loss rate for the X-ray O11 model. The radial trajectory of the outflow increases the radial optical depth, reducing the flaring of the photosphere. The increased vertical drag from the wind in the X-ray P19 model reduces the radial optical depth of the outflow, leading to a larger degree of flaring. The flaring in the EUV photosphere at radii $<10$~au arises because the dust-trap is more tightly radially confined in comparison to the other models (see fig. \ref{photdens} in appendix \ref{photapp}). In the EUV model, we see a thin wall of dust rising vertically at the inner edge of the disc before following approximately radial trajectories after $\sim3$~au, whilst the other models show a smoother, more radially extended density profile in the outflow for the first few au. This thin wall in the EUV model is less optically thick than the outflows at the inner edges of the X-ray models, meaning that the photosphere is located deeper in the disc. However, the outflow rapidly increases in radial optical depth as it moves out to $\sim10$~au, causing the photosphere to flare. Past 10~au, there is minimal difference in outflow morphologies between the X-ray and EUV models, and the photospheres exhibit similar levels of flaring.

In reference to criterion (ii) from section \ref{dispersalcriteria}, an interesting point to note is that the level of photospheric flaring is not dictated by shadowing; it is governed by the relative power of radiation pressure versus vertical drag due to wind, as this controls the structure of small grains in the disc surface. So, whilst the amount of flaring in the EUV and X-ray P19 models appears to potentially be a barrier to dust mass-loss from the disc, as dust can re-enter the disc if the trajectories do not have enough of a vertical component, it is actually set by the fact that the trajectories \textit{do} have enough of a vertical component. The inverse of this applies to the X-ray O11 model - although the photosphere is less flared up to $\sim30$ au, this is due to the radial trajectories of the grains, and we can see from Fig. \ref{2drunexample} that micron-sized grains have trajectories that drop below the photosphere, meaning they will not contribute to overall mass-loss from the disc. We explore the effect of this in our evolutionary calculations detailed in section \ref{evores}. 

\begin{figure*}
    \centering
    \includegraphics[width=0.99\textwidth]{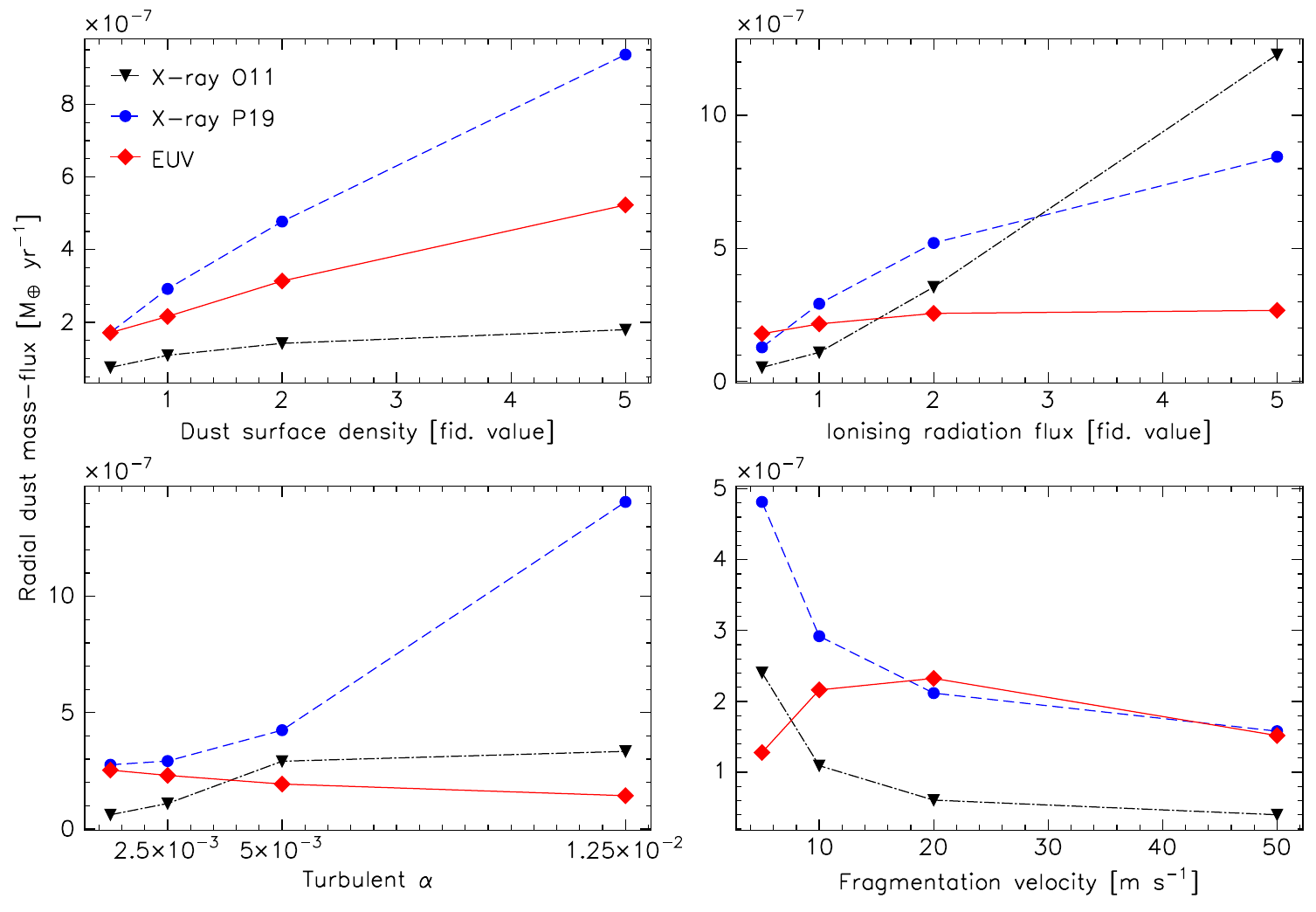}
    \caption{The dependence of the radial mass flux of dust above the dust-trap on different simulation parameters, namely the total dust surface density, the flux of ionising radiation that controls photoevaporation, the turbulent $\alpha$ and the fragmentation velocity. The results for each of the different wind models are shown for each parameter. Parameter values are normalised to those used in the fiducial simulations.}
    \label{dmldeps}
\end{figure*}
\subsection{Dust mass-fluxes}\label{dmfluxes}

To evaluate and compare the dust outflow for each simulation, we calculate the outwardly radial mass flux of dust above the interior of the disc at a radius of 1.1~$\times$~$R_\text{trap}$, with $R_\text{trap}$ being the radius at which the dust surface density in the trap is at a maximum. We choose a radius slightly past $R_\text{trap}$ as the flow is steadier here, meaning stabler mass fluxes can be calculated. 

Fig.~\ref{dmldeps} shows how the mass fluxes are affected by modulating each parameter. We see that for the fiducial models, the relative dust mass-fluxes are correlated with the total mass-loss rates for each wind model (given in Fig. \ref{windprofiles}). This is to be expected, given that larger mass-loss rates mean larger vertical gas velocities - this increases the drag on grains, increasing the supply of dust mass to the disc surface. When varying the dust surface density for all photoevaporation models, we find that increasing the amount of dust available leads to larger mass fluxes; this is because a larger surface density leads to a larger mass of small dust grains in the surface of the disc, increasing the mass reservoir available for acceleration. Notice, however, that the increase is sub-linear, and therefore, more massive discs will take longer to disperse. For the strength of ionising radiation, we find that all models show a correlation between increasing ionising flux and increasing mass fluxes. This can again be attributed to the increased wind velocity dragging a larger proportion of dust from the disc interior up to the surface of the disc. The mass-loss rates of the X-ray models have a similar dependence on X-ray luminosity that is stronger than the dependence of the EUV mass-loss rate on EUV photon flux; this is why the correlation is weaker for the EUV model compared to the X-ray models.

Both the turbulent $\alpha$ and fragmentation velocity affect the limiting maximum grain size set by fragmentation \citep{birnstiel2012},
\begin{equation}
    \label{afrag}
    a_\text{frag} \propto \frac{v_\text{frag}^2}{\alpha},
\end{equation}
but in opposite ways - the effect of this can clearly be seen in the correlations seen for the mass fluxes of the X-ray models with $\alpha$ and $v_\text{frag}$. A higher $\alpha$ or lower $v_\text{frag}$ decreases the maximum grain size, increasing the fraction of dust mass in small grains. For small grains, the balance between diffusive processes and radial forces due to gas drag is skewed in favour of diffusion, as they are well-coupled to the gas and thereby less affected by radial drift. With reduced maximum grain sizes, the total dust population is less concentrated in the dust-trap, spreading out over a wider radial region. This pushes the inner rim of the dust disc closer to the inner rim of the gas disc, which, in conjunction with the increased opacity of smaller grains, moves the mid-plane photosphere to smaller radii. This, in turn, decreases the mid-plane temperature at the dust-trap, which lowers the gas scale height, decreasing the gas density at the height of the photosphere. Reducing the gas density reduces the radial drag on grains that are being accelerated by radiation pressure, allowing large radial velocities for the grains. 

It is these large radial velocities that allow larger mass fluxes when moving to lower maximum grain sizes due to higher $\alpha$ or lower $v_\text{frag}$. The opposite effect is seen for decreasing $\alpha$ or increasing $v_\text{frag}$. One deviation from this is seen in the X-ray models is for the X-ray O11 model with the largest value of $\alpha$, where the viscous spreading is so strong that the dust-trap surface density falls to a value $7\times$ less than the fiducial model after 10,000 years, as dust is diffused out of the inner edge of the disc and into the wind. We do not see this in the X-ray P19 model as the trap is at larger radii and almost 10~au behind the edge of the gas disc, meaning the increase in viscous spreading does not force dust out of the inner edge of the disc, and the surface density does not reduce enough to limit mass-loss rates. The EUV model also diverges from the correlation seen in the X-ray models. As opposed to the mid-plane photosphere radius decreasing with lower maximum grain sizes, we find that it follows the opposite correlation, but to a weaker extent. This is because the disc inner rim in the EUV model is initially much closer to the star, and therefore the radiation pressure force on the grains is much larger ($f_\text{rad}\propto 1/r^2$). This especially affects bottom-heavy grain-size distributions (see fig. \ref{betas}) and overrides the diffusive spreading, pushing the dust, and therefore the photosphere, further out. These results imply that in the inner disc, radiation pressure dominates over the effects of gas drag and diffusion, whilst the opposite is true for large radii. \HL{In reference to our choice of fiducial $\alpha$ discussed in section \ref{setupICS}, we note that these results imply that using smaller values in the range \tenpow{-4}$-$\tenpow{-3} would not affect our qualitative results found in the evolutionary calculations described in section \ref{evores} for two reasons; that the effect of $\alpha$ on the photosphere plateaus as we move to lower values (as shown by Fig.~\ref{dmldeps}), and that dust is transported to the photosphere by advection from wind drag as opposed to turbulent diffusion.}

In summary, and in reference to the criteria set out in section \ref{dispersalcriteria}, we find that mass-loss rates from the trap are a few times less than the desired mass-loss rate of $10^{-6}$~\meyr for the fiducial models, however, more massive discs, stronger photoevaporation, more turbulence and lower fragmentation velocities can increase mass-loss, at least when the inner edge of the disc is at a few au. However, the relationship between mass-loss and disc mass is sub-linear, meaning that although mass-loss rates are larger, the disc will still take more time to clear. Likewise, larger turbulence will lead to larger mass of dust needing to be cleared, as the gas disc will evolve faster and therefore gap-opening will occur sooner. This also applies to larger photoevaporation rates. A similar issue may arise for lower fragmentation velocities, as these mean that more of the dust mass is in small grains, which drift slowly, leaving more dust mass in the outer disc at the onset of gap-opening. Stronger photoevaporation may also cause a separate issue by pushing the inner edge of the outer disc radially outwards at a faster rate, meaning that although there are large mass-losses via the outflow, the dust-trap will be refilled more quickly by the sweeping up of dust as the disc inner edge moves outwards.

We also find that the flaring of the disc, which in part controls whether dust can escape the disc without re-entering, is controlled not by shadowing due to the puffing-up of the inner rim, but by the morphology of the outflow. This is important for the evolution of the system, as the ability of dust to escape the disc can actually be hindered by the existence of the outflow itself.

\HL{A caveat to our results here is that our assumption of purely vertical gas velocities for the wind (see section \ref{PEwindsimp}) means that the vertical velocities of dust grains at the very inner rim are exaggerated and that there should also be an inwardly radial component. Reductions in the vertical velocities could lessen the effect of the dust outflow on the disc photosphere and reduce dust-mass loss rates. However, we do not think that this would alter our conclusions as the dust-trap (where the majority of the dust mass resides) is always located at a distance that is at least 10 gas scale heights behind the inner gas edge, and therefore several au behind, where the assumption of vertical gas velocities holds. Simulations that include the full 2D gas velocity solution should be performed in the future to investigate the effect on the results detailed here.}

\section{Results: evolutionary calculations}\label{evores}

For the full evolutionary calculations, we perform some additional steps after step (ii) in section \ref{setupICS} to improve the accuracy of the initial dust profiles, as here we do not only care about the steady-state dust mass-loss rates but also the time-dependent evolution. These steps are:

\begin{enumerate}    
    \item Given that the actual dust grain distribution has more complex features than the simple power-law profile of the MRN distribution \citep[see e.g.][]{birnstiel2011}, we perform a static, 1D evolution of the dust surface density distribution using a vertically integrated coagulation kernel \citep[see][]{birnstiel2010}, evolving the dust to coagulation-fragmentation equilibrium. We perform this calculation in 1D because, without dynamics, the dust would not settle, leading to erroneous results. As this calculation is static and therefore also doesn't account for the grain size limit caused by drift, we then apply exponential cut-offs to the distribution in regions of the disc where drift would limit the maximum grain size to below the fragmentation limit, i.e. $n_\text{cut}(a)=n(a)\exp\left[-(a/a_\text{d})^3\right]$, $a_\text{d}$ being the drift-limited maximum grain size. This then requires us to re-normalise the dust surface density in these regions. Finally, we then re-calculate the 2D dust density distribution using Eqn. \ref{2ddustdensity}.
    \item To relax the entire system before turning on the wind and radiation pressure, we run the dust dynamics, coagulation and temperature calculation for 1000 years.
\end{enumerate}

At this stage, we then ramp up the wind and radiation pressure up to their intended strengths over the course of 100 years. From this point on, the simulations are run until we are able to establish the efficacy of dust disc dispersal via radiation pressure outflows; typically a timescale of $0.5-1$~Myr.

To evaluate the effect of radiation pressure over the full disc evolution across a range of possible disc configurations, we ran a parameter study in which we varied the stellar type, the dust fragmentation velocity ($v_\text{frag}$) and the photoevaporation model. The values of these parameters and others used in the suite of simulations can be seen in Table \ref{tab_params}. We also varied the strength of photoevaporation in each model; the X-ray luminosity ($L_X$) for the X-ray models and the EUV ionising flux ($\Phi$) for the EUV models. Two-grain internal densities can be seen in the table - one was used for the majority of our study as it comes from the DSHARP \citep{DSHARP2018} opacity model, whilst the other density was used for simulations where the \cite{ricci2010} opacity model was applied. 

\subsection{Fiducial model}

\begin{figure}
    \centering
    \includegraphics[width=0.99\columnwidth]{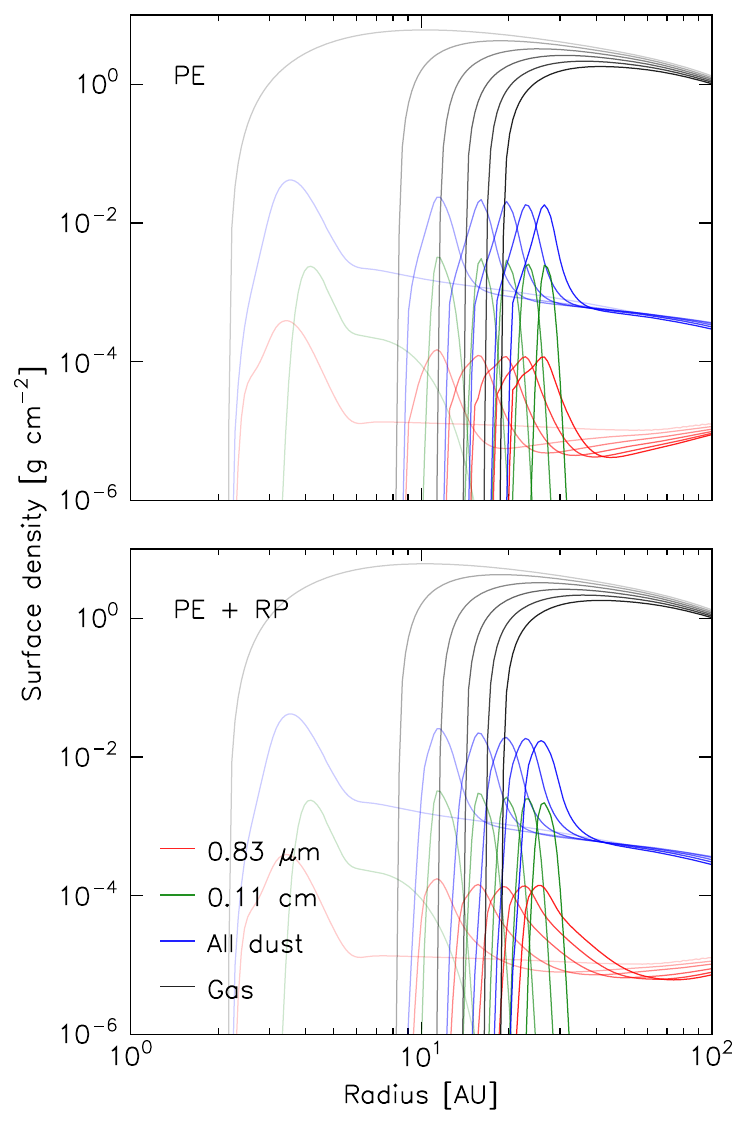}
    \caption{The evolution of dust and gas surface density profiles for the fiducial X-ray O11 configuration with just photoevaporative (PE) effects (top panel) and with both PE and radiation pressure (RP) effects (bottom panel). Two different grain sizes are shown, representing small and large grains. For this simulation, the gas gap opened after $\sim3.6$~Myr and the inner dust disc fully drained after another $\sim 50$~kyr. Starting at the end of inner disc drainage, this figure shows time snapshots at 0~year and then at intervals of 0.2~Myr, ending at 1~Myr of evolution. The line opacity increases with time.} 
    \label{JOfidsigmas}
\end{figure}

\begin{figure*}
    \centering
    \includegraphics[width=0.99\textwidth]{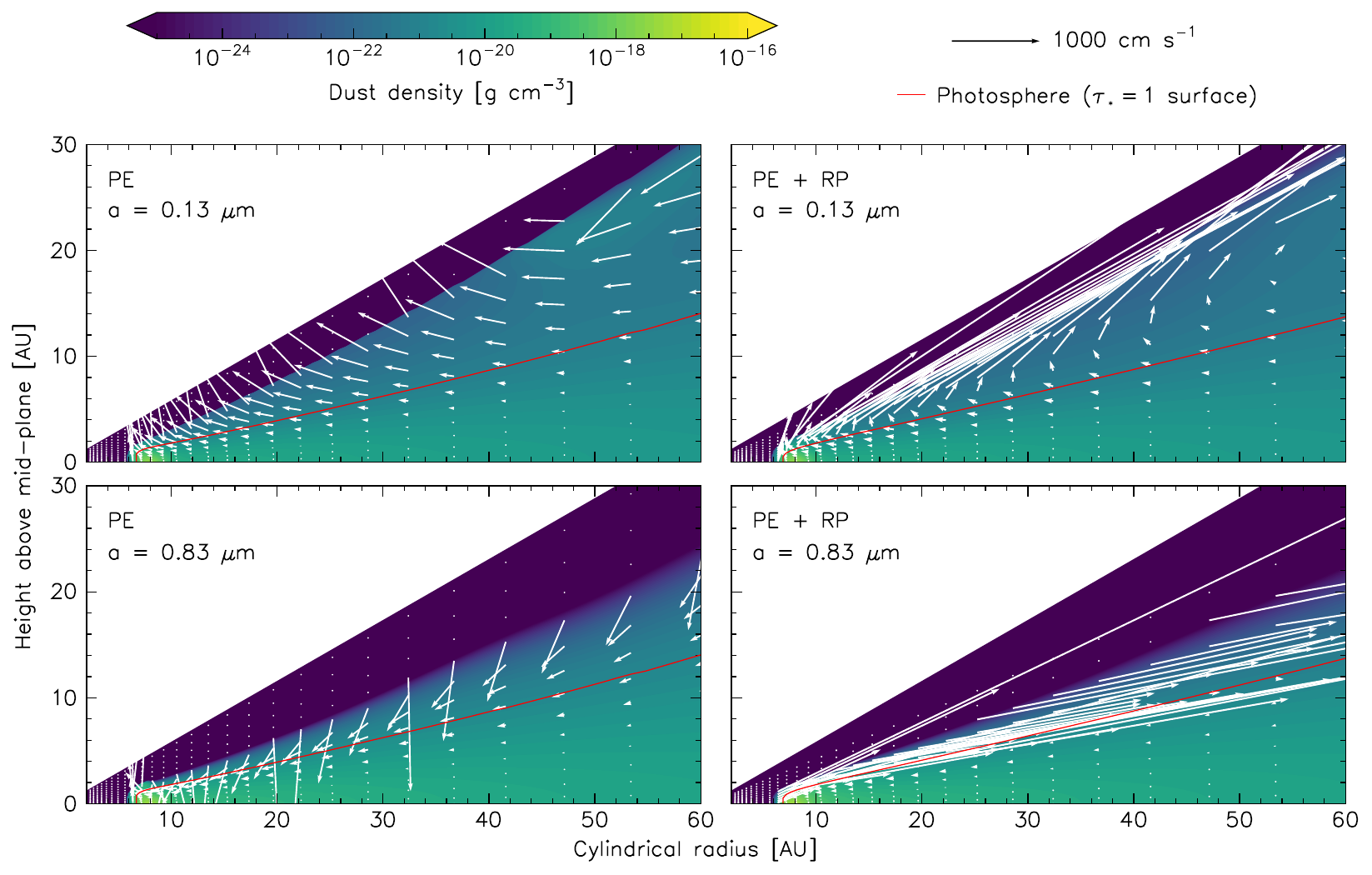}
    \caption{2D density and velocity profiles for two different grain sizes in the fiducial X-ray O11 model. The left and right columns show simulations with only photoevaporative (PE) effects and with both PE and radiation pressure (RP) effects respectively. The top and bottom rows show the profiles for 0.13~\&~0.83~\unit{\micron} sized grains respectively. The photosphere, defined as the surface at which the optical depth to stellar radiation, $\tau_*$, is unity, is shown in red.  This particular snapshot is after 75,000~years of evolution.}
    \label{JOfid2Ds}
\end{figure*}

The X-ray O11 model was used in the initial study of disc dispersal under the influence of radiation pressure \citep{owen2019}. For this photoevaporation model, we define our fiducial simulation as using the 0.7~\msun star and having the parameters $\log_{10}{L_X}=30$ and $v_\text{frag}=10$~\unit{\metre\per\second}. Fig.~\ref{JOfidsigmas} shows the evolution of the surface density profiles of both gas and dust for the fiducial model with and without the effects of radiation pressure. These surface densities were calculated from the 2D simulations by integrating over the vertical dimension. For the individual grain sizes, vertically integrated, mass-grid-independent surface densities were calculated from the volume densities via

\begin{equation}
    \label{dustsigma}
    \sigma_d(m) = \int_{-\infty}^{\infty} m \frac{\partial \rho_d}{\partial m} dz.
\end{equation}

We can see that the inclusion of radiation pressure does not lead to a significant change in the evolution of the dust surface density. For both 2D simulations, the initial dust mass was 1.93~$M_\oplus$ and decreased to $\sim1.82~M_\oplus$ after 1~Myr, giving average mass-loss rates of just above \tenpow{-7}~\meyr, which agrees with the radial mass-loss rates calculated for the fiducial model in section \ref{dmfluxes}. Whilst this mass-loss rate is less than the desired \tenpow{-6}~\meyr, we would still expect to see a more significant reduction in the dust-trap surface density over the course of 1~Myr, however, the mass in the dust-trap (defined as the region between the inner edge of the disc and the `knee' in the dust surface density profile, which is at around 6 au for the first time snapshot in Fig. \ref{JOfidsigmas}) actually increases from 0.11 to 0.55~\mearth over the course of the photoevaporation and radiation pressure simulation. This evidently does not fit with observational constraints of removing the observational disc component within $\sim10\%$ of the disc lifetime, and implies that the impact of inward dust drift and the outward motion of the disc's inner edge due to photoevaporation refilling the dust-trap means that the required mass-loss rates need to be higher. For this fiducial model, the loss rate would have to be about at least five times higher to overcome the refilling, closer to the estimated desired value of \tenpow{-6}~\meyr. The only significant difference visible in the surface density profiles is in the small grain population, where the inclusion of radiation pressure leads to a shallower and less steep decrease in density at radii exterior to the trap. 

\begin{figure}
    \centering
    \includegraphics[width=0.99\columnwidth]{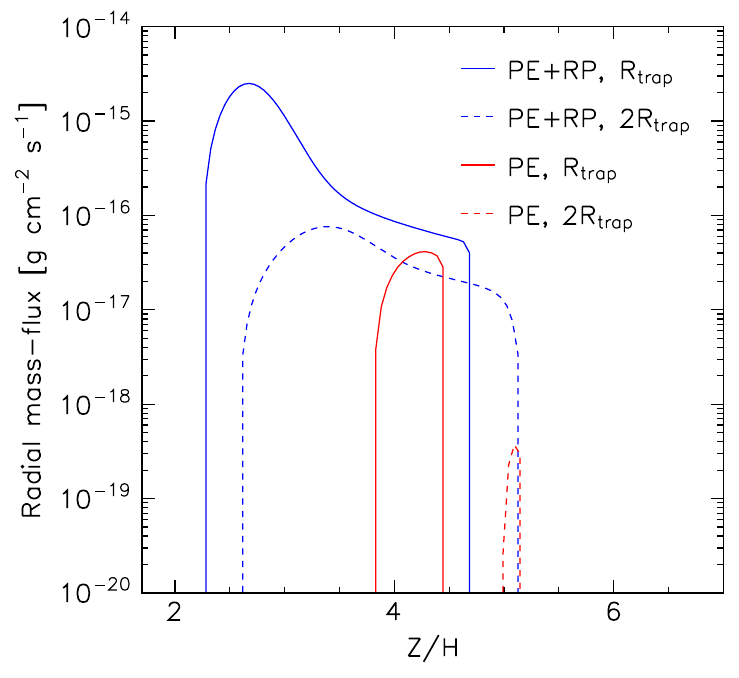}
    \caption{The spherically-radial mass flux of grains up to 2~\unit{\micron} in size for the fiducial X-ray O11 configuration. The $x$-axis is plotted in units of $Z/H$, $H$ being the gas scale height set by the respective mid-plane temperature for each simulation and location. The mass fluxes are shown for both the model with just photoevaporative (PE) effects and the model with  PE and radiation pressure (RP) effects. The mass fluxes are shown at the mid-plane radius of the dust-trap, $R_\text{trap}$, and $2\times$ this value. This snapshot is after 75,000 years of evolution.} 
    \label{JOradmassflux}
\end{figure}

If we compare the 2D density and velocity structures for small grains, we can investigate why the mass-loss rates are not high enough. Fig.~\ref{JOfid2Ds} shows such profiles for $\sim0.1~\&~\sim1~\mu$m grains. In the purely photoevaporative model, we see the sub-micron-sized dust grains well-coupled to the gas flow; they move radially inwards due to viscosity and vertically \HL{upward}s due to the photoevaporative wind flow. Larger, micron-sized grains are less well-coupled to the gas, moving vertically \HL{upward}s only at the very inner edge of the disc where the vertical mass flux of the photoevaporative wind is largest. In contrast, in the photoevaporation + radiation pressure model, sub-micron-sized grains are driven along radial trajectories due to radiation pressure. As discussed in section \ref{outflowcharacteristics}, we see that the micron-sized grains are also impacted strongly by radiation pressure, however their trajectories are not away from the disc, but deeper into the disc interior. We can see the effect of this by looking at the spherically-radial mass flux of grains up to 2~\unit{\micron} in size through different regions of the disc. These can be seen for both the photoevaporation model and photoevaporation + radiation pressure model at the location of the dust-trap and $2\times$ this radius in Fig.~\ref{JOradmassflux}. The sharp cut-offs at low and high heights are caused by the transition from inward radial drift (negative mass flux) to outward radial motion caused by photoevaporation/radiation pressure and the flooring of dust that travels above the X-ray wind base. Whilst radial mass fluxes for the photoevaporation + radiation pressure model are high at the trap location (comparable to those found in the simulations of \citealt{owen2019}), they reduce by several orders of magnitude further out in the disc. This is due to the re-interception of the disc material decelerating the dust via drag. The dust can then drift back into the pressure trap as opposed to escaping. Linking back to criterion (ii) from section \ref{dispersalcriteria}, this re-interception severely limits dust removal from the disc by radiation-pressure-driven outflows, regardless of large radial mass fluxes at the trap location. 

\begin{figure}
    \centering
    \includegraphics[width=.99\columnwidth]{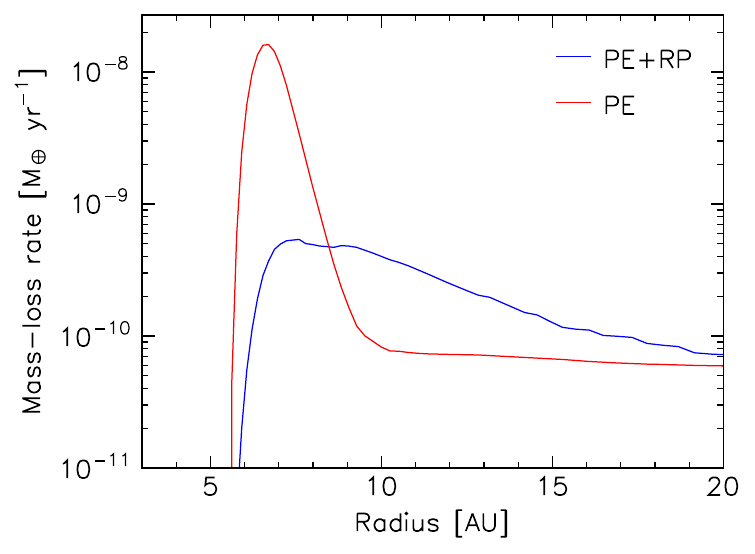}
    \caption{The total mass-loss rate of dust through the wind base for the fiducial X-ray O11 simulation, for both the model with just photoevaporative (PE) effects and the model with  PE and radiation pressure (RP) effects. This snapshot is after 75,000 years of evolution.}
    \label{windbaseml}
\end{figure}

We can also see the difference caused by the inclusion of radiation pressure by looking at the mass-loss through the wind base, shown for the two cases with and without radiation pressure in Fig. \ref{windbaseml}. Without radiation pressure, the wind drags a large mass of dust vertically \HL{upward}s at the very inner edge of the disc, whilst with radiation pressure we see that the additional radial force decreases the flux of dust crossing the wind base, an effect suggested by \cite{booth2021} in their work on dust entrainment. This means that whilst the addition of radiation pressure increases the mass of dust that is pushed radially outwards to further radii, it actually decreases the loss of dust to the wind at the disc inner rim. 

\subsection{Parameter study - mass-loss rates and final masses}

In this section, we study whether any regions of parameter space can allow dust to be removed efficiently on secular timescales with the evolution of both the dust and gas in the disc. As previously found in section \ref{dmfluxes}, the regions of parameter space that lead to the largest mass-loss rates may also lead to increased disc masses, and therefore longer dispersal times. 

\subsubsection{X-ray O11 model}

\begin{figure}
    \centering
    \includegraphics[width=0.99\columnwidth]{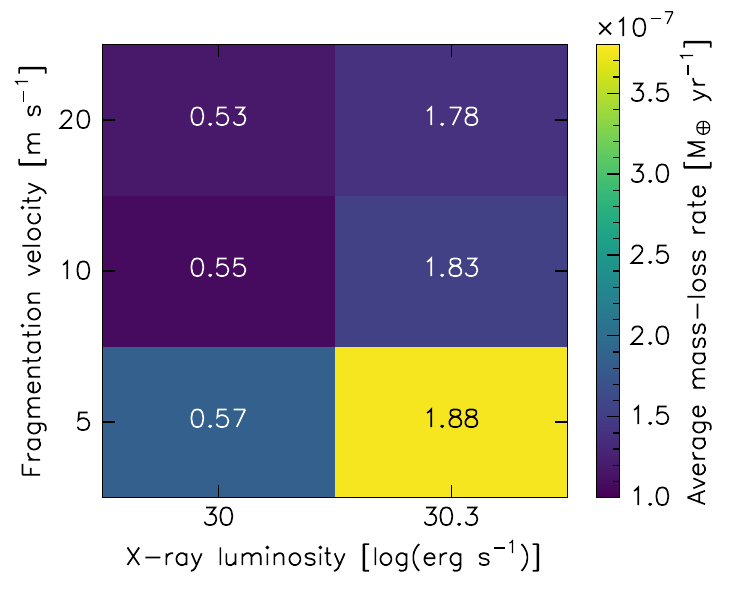}
    \caption{The average mass-loss rates and final dust-trap masses for the X-ray O11 parameter study with the 0.7~\msun star. The colour-map indicates the mass-loss rates whilst the numbers in each box indicate the final dust masses, 1~Myr after gap-opening.} 
    \label{joparamstudy}
\end{figure}

Fig.~\ref{joparamstudy} shows the average mass-loss rates and final dust-trap masses 1~Myr after \HL{the time of} gap-opening for the X-ray O11 parameter study with the 0.7~\msun star. The ages of the discs at gap-opening were 3.6~\&~2.1~Myr for X-ray luminosities of 30~\&~30.3, respectively. As with the fiducial model, in all of the runs, the mass in the dust-trap increased over the course of the simulation. This means that none of the parameter choices leads to the removal of the dust component in the timescales required by observational constraints, as the dust trap remains too massive and, therefore optically thick, even 1~Myr after gap-opening.  

As suspected in section \ref{dmfluxes}, we find that lower fragmentation velocities and higher X-ray luminosities lead to larger mass-loss rates but also mean that there is a larger mass in the dust disc at the onset of gap-opening. This results in the low fragmentation velocity, high X-ray luminosity case having the largest dust mass remaining 1~Myr after gap-opening even though the mass-loss rate is largest. 

Runs with the 0.4~\msun star and a fragmentation velocity of 10~\unit{\metre\per\second} produced dust mass-loss rates of \num{4.06e-7}~\&~\num{1.16e-6}~$M_\oplus$~yr\minone for X-ray luminosities of 30~\&~30.3 respectively. The respective disc ages at \HL{the time of} gap-opening were 2.9~\&~1.6~Myr for these luminosities. Whilst the mass-loss rates are larger than the rates for the 0.7~\msun star, the discs around the 0.4~\msun star also had more dust present at the onset of gap-opening, so 1~Myr after gap-opening, the simulations had 0.75~\&~2.2~$M_\oplus$ of dust remaining in the trap. Attempting simulations for the 0.4~\msun star at fragmentation velocities of 5~\unit{\metre\per\second}, we found that at the end of the 1D simulations the dust-to-gas ratio at the inner edge of the outer disc was greater than unity. This is because the smaller dust grains, which are less well-coupled to the gas, are less affected by the outwardly radial force provided by the super-Keplerian gas, and so are able to build up closer to the inner rim as photoevaporation removes gas and increases the radius of the cavity. This feature was also observed in the work by \cite{alexander07}. In this regime, the dust would start to act independently from the gas and would create strong back-reaction effects on the gas, none of which are included in our model, so we discarded these simulations and did not run 2D simulations. However, our results for the 0.7~\msun star indicate that discs with these parameters would also not clear on fast enough timescales. 

\subsubsection{X-ray P19 model}

As with the X-ray O11 model, we ran simulations with and without the effect of radiation pressure for the X-ray P19 model with the 0.7~\msun star and the parameters $\log_{10}{L_X}=30.3$ and $v_\text{frag}=5$~\unit{\metre\per\second}. The gap opened after 1.5~Myr for this configuration. We find that radiation pressure again does not aid dust removal much. 0.25~Myr after gap-opening, the total dust masses remaining in the dust traps were 0.28~\&~0.23~$M_\oplus$ for the model with just photoevaporation and with both photoevaporative and radiation pressure effects, respectively. Given that this mass of dust still constitutes an optically thick trap, we cannot meet the observational constraints. In contrast to the X-ray O11 model, our 2D simulations end $\sim0.3$~Myr after gap-opening because the gas mass-loss rates are higher for the X-ray P19 model (as can be seen in Fig.~\ref{windprofiles}), leading to the inner rim of the disc exiting our grid boundaries in less than 0.5~Myr. 

\begin{figure}
    \centering
    \includegraphics[width=0.99\columnwidth]{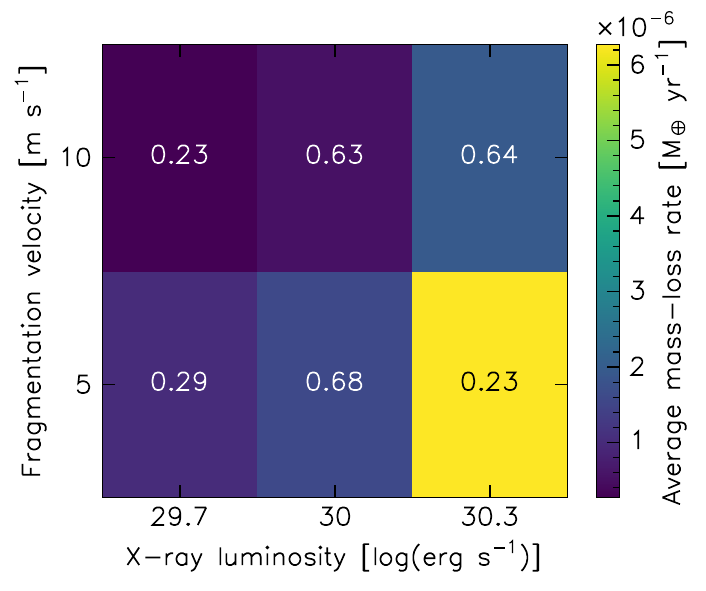}
    \caption{The average mass-loss rates and final dust-trap masses for the X-ray P19 parameter study with the 0.7~\msun star. The colour map indicates the mass-loss rates whilst the numbers in each box indicate the final dust masses, 0.25~Myr after gap-opening.} 
    \label{Picparamstudy}
\end{figure}

Fig.~\ref{Picparamstudy} shows the average dust mass-loss rates and final dust-trap masses 0.25~Myr after gap-opening for the X-ray P19 parameter study with the 0.7~\msun star. The ages of the discs at \HL{the time of }gap-opening were 3.3,~2.1~\&~1.5~Myr for X-ray luminosities of 29.7,~30~\&~30.3 respectively. As with the X-ray O11 model, none of the parameter choices reduce the dust-trap mass enough to meet observational constraints. We did not include runs with fragmentation velocities of 20~\unit{\metre\per\second} as these led to the lowest mass-loss rates for the X-ray O11 model. The trend in mass-loss rates is the same as for the X-ray O11 models, with higher mass-loss for lower fragmentation velocities and higher X-ray luminosities. For final dust-trap masses, we see a similar trend of an increase in final mass for larger X-ray luminosity for a fragmentation velocity of 10~\unit{\metre\per\second}. However, for a fragmentation velocity of 5~\unit{\metre\per\second} we find that the final mass actually decreases when moving from an X-ray luminosity of 30 to 30.3. This is because, although the higher luminosity simulation starts with more mass due to shorter disc ages at \HL{the time of }gap-opening, the mass-loss rate is high enough that in the same time period, it can remove enough mass to give a final mass lower than the lower luminosity simulation. 

For the 0.4~\msun star, the same trend as in X-ray O11 was found. Whilst the mass-loss rates were larger than their 0.7~\msun star counterparts (all in the range \num{5.7e-7}-\num{1e-5}~$M_\oplus$~yr\minone), the final masses were also all larger (e.g. 1~$M_\oplus$ for the 5~\unit{\metre\per\second}, 30.3 simulation; this with a mass-loss rate of \num{1e-5}~$M_\oplus$~yr\minone). This is because the ages of the discs at \HL{the time of }gap-opening were shorter than their 0.7~\msun star counterparts; 2.9,~1.9~\&~1.4~Myr for X-ray luminosities of 29.7,~30~\&~30.3 respectively.

We also ran models with the X-ray P19 wind model but using the \cite{ricci2010} opacities as opposed to the DSHARP opacities, however this change only slightly affected the quantitative results, with therefore no real effect on the qualitative outcomes of the simulations.
% \begin{figure}
%     \centering
%     \includegraphics[width=0.99\columnwidth]{figures/Picsigmas_fid.pdf}
%     \caption{The evolution of dust and gas surface density profiles for the X-ray P19 model with just photoevaporative (PE) effects (top panel) and with both PE and radiation pressure (RP) effects (bottom panel). The parameters used for these simulations were $\log_{10}{L_X}=30.3$ and $v_\text{frag}=5$~\unit{\metre\per\second}. Two different grain sizes are shown, representing small and large grains. For this simulation, the gas gap opened after $\sim1.5$~Myr and the inner dust disc fully drained after another $\sim 120$~kyr. Starting at the end of inner disc drainage, this figure shows time snapshots at 0~year and then at intervals of 100~kyr, ending at 300~kyr of evolution. The line opacity increases with time.} 
%     \label{Picfidsigmas}
% \end{figure}

\subsubsection{EUV model}

For the EUV model, we used values of photon ionising flux of \num{1e43}~\&~\num{5e43}~photons~s\minone. For the 0.7~\msun star, \HL{the ages of the disc at the time of gap-opening} were 12~\&~6.3~Myr for \num{1e43}~\&~\num{5e43}~photons~s\minone
respectively, and for the 0.4~\msun star the low flux simulation had an age $>20$~Myr, meaning we discarded it for being too long to match observations, and the high flux simulation had an age of 6.9~Myr. We were also only able to include simulations with fragmentation velocities of 10~\unit{\metre\per\second}, as simulations with 5~\unit{\metre\per\second} led to dust-to-gas ratios greater than unity at the inner rim of the disc.

The EUV models were the only models where observational constraints on dispersal were met, however this was not solely due to the addition of radiation pressure. For the high flux, 0.7~\msun star simulation with just photoevaporation, the mass-loss rate was \num{2.49e-6}~$M_\oplus$~yr\minone and had 0.18~$M_\oplus$ of dust remaining in the trap after 0.3~Myr, whilst the simulation with radiation pressure had respective values of \num{2.55e-6}~$M_\oplus$~yr\minone and had 0.12~$M_\oplus$. The simulation with radiation pressure having a final dust-trap mass that is around a third less than the final mass without radiation pressure, whilst only a marginally larger mass-loss rate, indicates that radiation pressure causes more mass to be moved to larger radii without actually leaving the disc system. Again, we choose a time cut-off due to the disc's inner radius leaving the simulation domain shortly after 0.35~Myr. For the low flux, 0.7~\msun star simulation with radiation pressure, we find that the dust is totally removed whilst the disc inner rim is within the grid, dropping to less than \num{5e-3}~$M_\oplus$ after 0.8~Myr; an average mass-loss rate of \num{5.47e-7}~$M_\oplus$~yr\minone. For the 0.4~\msun star, the high flux simulation dropped to 0.1~$M_\oplus$ after 0.3~Myr with an average mass-loss rate of \num{3.14e-6}~$M_\oplus$~yr\minone.

These mass-loss rates are comparable to the rates found for the other two models; one reason why the discs can be more successfully cleared is that the initial dust masses at \HL{the time of }gap-opening are almost an order of magnitude lower than the initial masses for the other models. \HL{Higher photon fluxes would decrease the lifetime of the disc pre-gap-opening; however, we are already using atypically large fluxes - anything larger would be very unlikely in real systems.} Also, importantly, the EUV surface-density loss rates are much larger in the direct rather than the diffuse regime; this can be seen in the fact that the integrated fiducial EUV gas mass-loss rate is comparable to the fiducial X-ray mass-loss rates (see Fig. \ref{windprofiles}) in the direct regime, however, the pre-gap-opening disc lifetime is several times longer. This allows more gas and dust to be lost onto the star before outer disc dispersal commences.

\begin{figure*}
    \centering
    \includegraphics[width=0.99\textwidth]{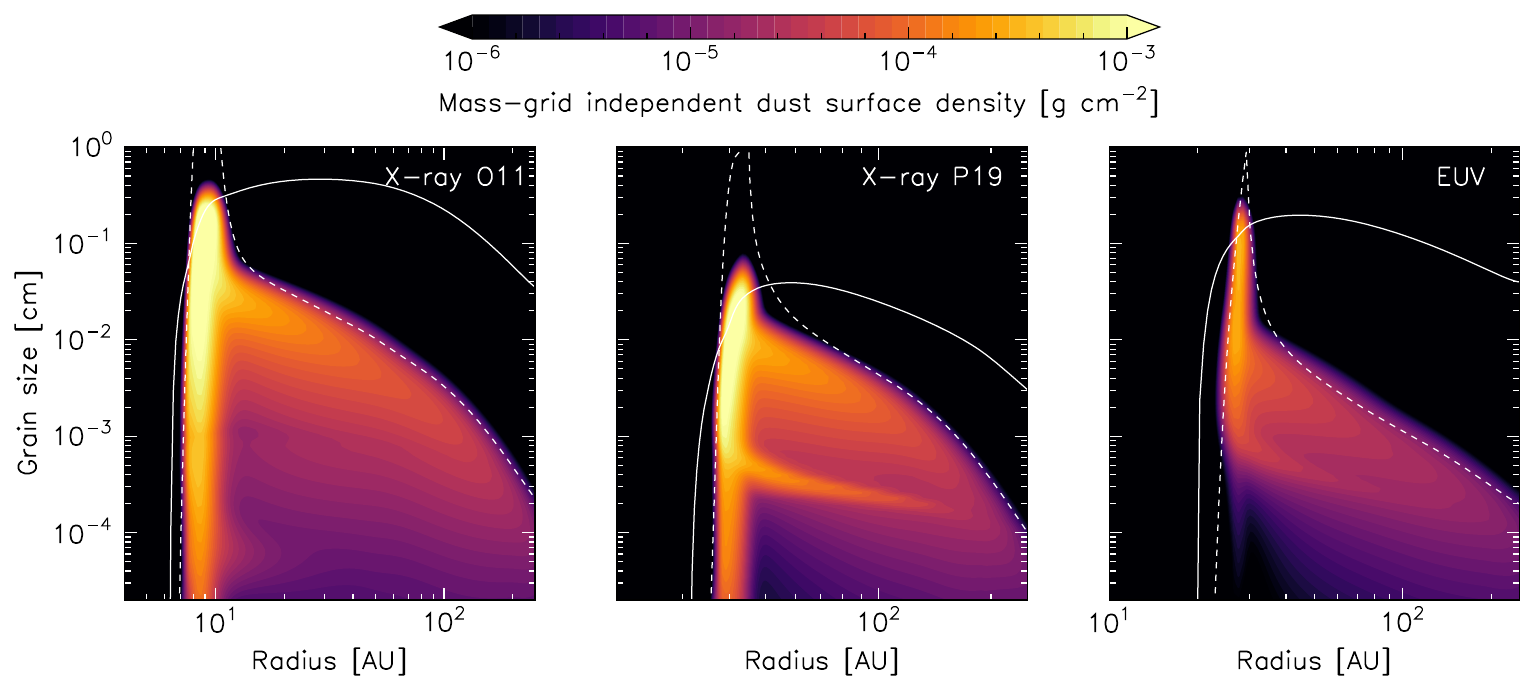}
    \caption{The vertically-integrated grain-size distribution with respect to radius for each of the three wind models. The parameters for each of the models are; X-ray O11: $\log_{10}{L_X}=30$, $v_\text{frag}=10$~\unit{\m\per\second}, X-ray P19: $\log_{10}{L_X}=30.3$, $v_\text{frag}=5$~\unit{\m\per\second}, EUV: $\Phi=\num{5e43}$~photons~s\minone, $v_\text{frag}=10$~\unit{\m\per\second}, all with the 0.7~\msun star. For reference, the approximate maximum limits on grain size due to fragmentation and drift (given by \citealt{birnstiel2012}) are over-plotted in solid white and dashed white respectively. These particular snapshots are after 100,000~years of evolution.}
    \label{graindistcomp}
\end{figure*}

\subsubsection{Evolution summary}

In summary, the evolutionary calculations demonstrate that only low mass discs ($\lesssim$~1~\mearth) can be efficiently dispersed via the action of radiation pressure and photoevaporative winds. For each of the photoevaporative models we find mass-loss rates of between \tenpow{-7}-\tenpow{-6}~\meyr, however only in EUV models \HL{with atypically large high-energy photon fluxes} do we find that the disc can be dispersed sufficiently quickly. Calculating a representative dispersal timescale from the ratio of the initial dust disc mass to the mass-loss rate for each simulation can give us a metric to understand when disc dispersal is efficient. For the successful EUV simulations, this timescale is $<0.5$~Myr, whilst the X-ray P19 and O11 simulations have timescales of order 1~Myr and 10~Myr, respectively. The X-ray P19 high photoevaporation rate and low fragmentation velocity run has a timescale of 0.6~Myr, making it the most efficiently dispersing disc from the X-ray runs. We also find that radiation pressure competes with entrainment in the wind and decreases the dust mass reaching the wind base at the inner rim of the disc by pushing material out to larger radii.

In the majority of cases, the refilling of the dust trap through radial drift and the outward motion of the disc's inner rim means that the required mass-loss rates from the trap need to be larger by up to an order of magnitude. Looking at our results in comparison to the criteria for disc dispersal set out in section \ref{dispersalcriteria}, this means that the discs are not fulfilling the first criterion, that the mass-loss out of the trap is sufficiently large. This is despite fulfilling the third criterion, that the mass-loss is not limited by fragmentation at the mid-plane or transportation of small grains above the photosphere, as the mid-plane densities of the large, fragmenting dust grains in each simulation are of the order a few~\tenpow{-15}~\unit{\gram\per\cm\cubed}, greater than the critical value calculated in section \ref{dispersalcriteria}. We can also see from fig. \ref{graindistcomp}, the dust grain-size distributions for simulations from each wind model, that both the X-ray models show an abundance of small grains at the trap, indicating they are not over-depleted by the outflow. 
% However, in the case of some of the EUV models, the mass-loss rate does appear to be fragmentation limited given the depletion of small grains at the mid-plane (as seen in Fig.~\ref{graindistcomp}), but, for this simulation, this limitation does not prevent the mass-loss rate from being large enough to clear the trap of dust.
\begin{figure}
    \centering
    \includegraphics[width=0.99\columnwidth]{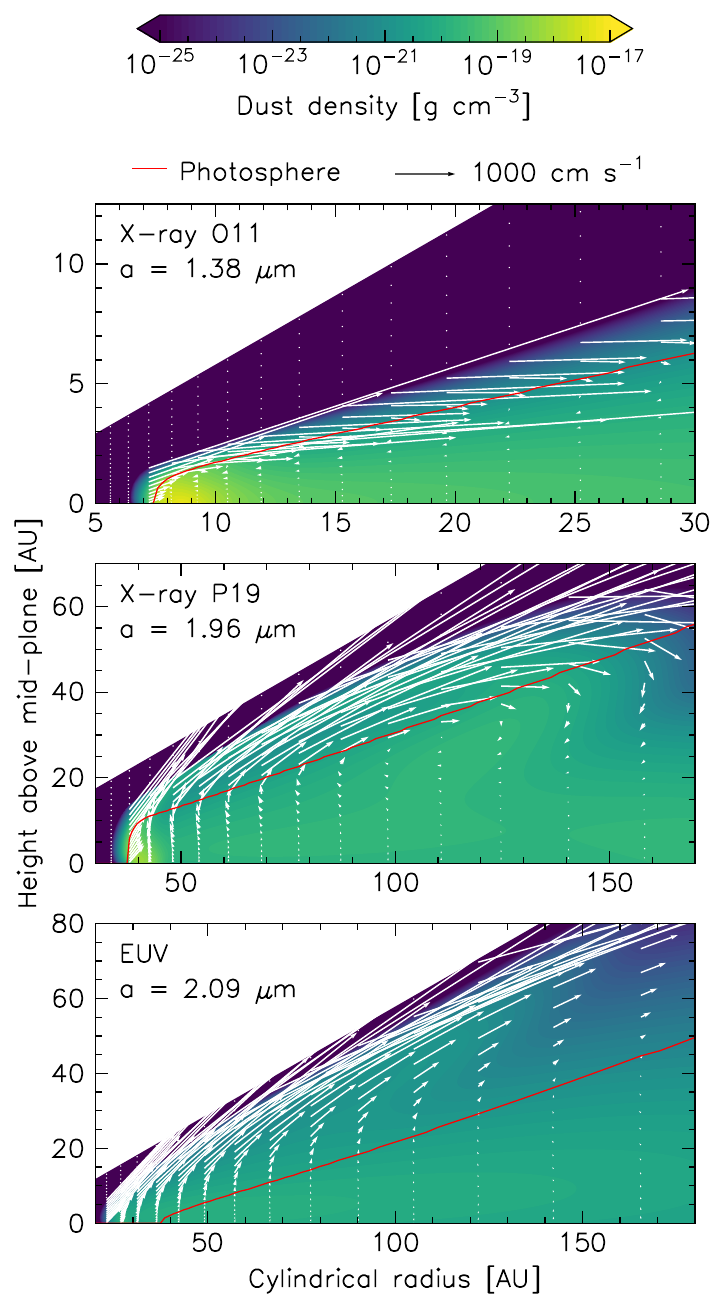}
    \caption{The outflow of micron-sized grains formed for each of the simulations shown in fig. \ref{graindistcomp}. All models are shown at a snapshot time of 100,000 years.}
    \label{2Doutflow}
\end{figure}

Another issue that limits the mass-loss rates relates to the second criterion, which is that the dust leaving the trap must be able to escape the disc. As discussed earlier, for the X-ray O11 model, from Fig.~\ref{JOradmassflux}, we can see that the radial mass flux at the radius of the trap and at 2$\times$ this radius decreases by at least an order of magnitude. This is because the trajectories of the dust grains above $\sim$~\unit{\micron}, which would make up a large proportion of the mass in the outflow, point at an angle above the mid-plane that is below the angle above the mid-plane made by the photosphere of the disc. This leads to radiation pressure exerting a weaker force on the grains and also increased drag with the gas in the disc. Ultimately, these grains then re-enter the disc interior, settling downwards and drifting inwards towards the disc's inner edge. This behaviour is consistent across all of the X-ray O11 simulations. Fig.~\ref{2Doutflow} shows these trajectories for micron-sized grains for the fiducial X-ray O11 simulation, alongside similar trajectories for the simulations made with the other wind models shown in Fig.~\ref{graindistcomp}. We see that dust in the X-ray P19 simulation is launched more effectively than in the X-ray O11 model, with the trajectories at the very inner edge pointing out of the disc and through the wind base, but that the decrease in wind drag with radius causes dust to fall back into the disc at larger radii. This again leads to refilling of the dust-trap as the dust drifts back inwards and the trap location moves outwards due to photoevaporation. The dust is, therefore, essentially cycled through the disc, moving from the trap to larger radii before drifting in to repeat the cycle. In the EUV models, we see micron-sized grain outflows that are able to escape the disc by crossing the wind base at the inner edge of the disc and at larger radii. The combination of a stronger wind and a decreased initial dust-trap mass leads to a low density, vertically distended small grain dust distribution and therefore a low-altitude photosphere. This allows more dust to experience strong forces due to radiation pressure, through which an efficient outflow can form.

Although one could infer from our results that only an EUV wind driven by atypically large high-energy photon fluxes could explain the observations, we do not believe that this is a satisfying conclusion. What is clear, is that the shape of the wind mass-loss profile with radius is very important, along with the mass of dust in the disc at \HL{the time of }gap-opening. \HL{Smoother mass-loss profiles that are less peaked at the inner rim (i.e. EUV-like profiles) aid in preventing infall of dust at larger radii.} Better constraints on the shape of X-ray wind profiles could allow micron-sized grains to escape the disc in the manner found for EUV winds. \HL{However, recent hydrodynamic and thermochemical simulations of X-ray photoevaporation by \cite{sellek2024} imply that this is unlikely, as they include additional cooling pathways to those found in the models of O11 and P19 and find significantly lower temperatures in the wind (driven predominantly by atomic cooling through neutral collisions), leading to lower gas mass-loss rates; real systems therefore may have even lower dust mass-loss rates than we have found in this work.} 

\section{Dust grain-size distribution: `levitating' grains}\label{graindist}

In addition to mass loss, the inclusion of the vertical dimension in our dust growth and dynamics simulations allows us to study the effect of winds and radiation pressure on the dust population in both grain-size distribution and spatial distribution; we have, in fact, already seen that the flaring surface of the disc is dictated by the outflow. Fig.~\ref{graindistcomp} shows the dust grain-size distribution as a function of radius for three simulations after \tenpow{5}~years, one for each wind model. As with Fig.~\ref{JOfidsigmas}, we plot the vertically-integrated, mass-grid-independent dust surface density (see eqn. \ref{dustsigma}) for each grain size. The grain-size distributions show that, as expected, the maximum grain size is set by fragmentation in the region of the dust-trap at the inner edge of the disc, whilst set by dust drift in the outer parts of the disc. 

Another notable feature is the enhancement of micron-sized grains at radii exterior to the trap seen in all three wind models, most prominently the X-ray P19 model. This enhancement arises because grains with sizes of around $1-$a~few~\unit{\micron} are `levitated' by the balance between drag via wind and gravitational settling. Grains much larger than this are too massive and have negative vertical velocities at all heights, whilst much smaller grains have positive vertical velocities at all heights; they are light enough to be dragged up above the photosphere and out of the disc. The exact size of the levitating grains depends on the wind strength, demonstrated by the fact that the enhancement is seen at larger grain sizes for the X-ray P19 and EUV wind models, which have greater wind loss rates in these examples. 

\begin{figure}
    \centering
    \includegraphics[width=0.99\columnwidth]{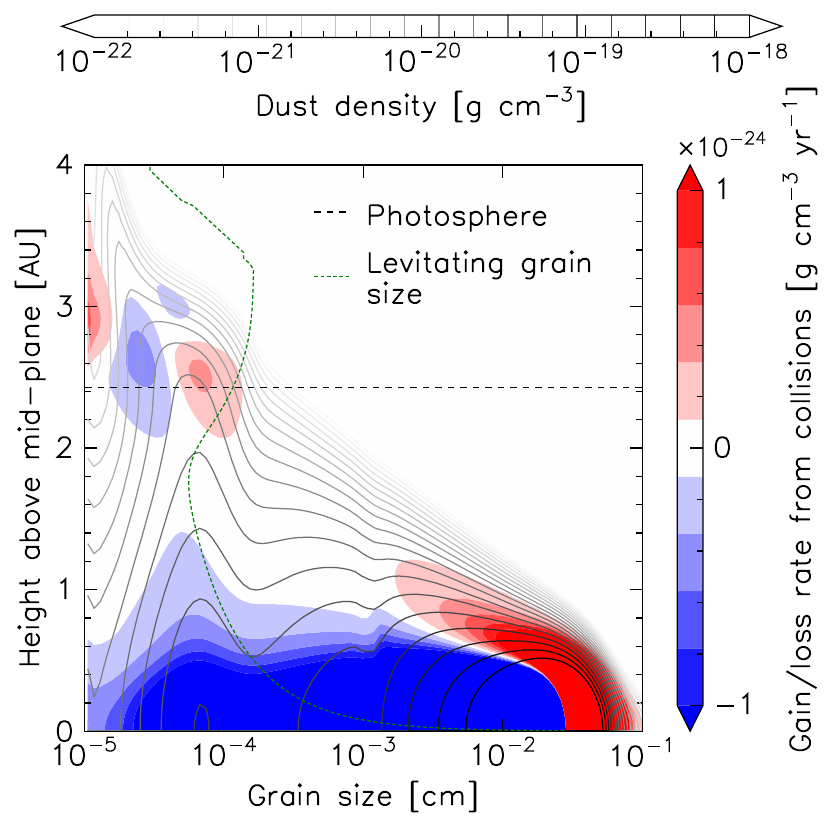}
    \caption{The dust grain-size distribution with respect to height (grey contours) after \tenpow{5}~years at a radius of 13 au, behind the dust-trap, for the fiducial X-ray O11 simulation. The red/blue, filled contours show the overall production/destruction rates of grains due to coagulation and fragmentation. The photosphere, defined as the surface at which the optical depth to stellar radiation, $\tau_*$, is unity, is shown by the black dashed line, and the levitating grain size $\left(v_Z(Z,a) = 0\right)$ is shown by the green dotted line.}
    \label{graindist_ratesbehindtrap}
\end{figure}

We can gain insight into why the density of these grains is enhanced by looking at the relationship between dust dynamics and growth in the vertical dimension. Fig.~\ref{graindist_ratesbehindtrap} shows the vertical dust grain-size distribution at a radius behind the dust-trap for the fiducial X-ray O11 simulation. In the dust density contours, we can see the over-density of grains at around 0.7~\unit{\micron} that extends to the mid-plane. 

The coagulation rates show that small grains grow into large grains around the mid-plane; they are not replenished, however, because, in this region of the disc, large grains drift radially inwards before they can grow to sizes larger than the fragmentation threshold. At intermediate heights (between $\sim1-2$~au), the distribution is in equilibrium, where coagulation rates are low because the relative velocities between grains are small. However, as we reach the upper layers of the disc, we see that small grains being dragged up from the mid-plane by the wind are coagulating into levitating, micron-sized grains. This is because their relative laminar velocities rapidly increase with height due to drag from the now high-velocity wind and radial acceleration from radiation pressure, especially above the photosphere, increasing collision rates (see Fig. \ref{velocitymags} in appendix \ref{levitatinggrainapp}). 

However, as the grains grow into these levitating grains, their growth rates drop as the relative laminar velocities between themselves and similar-sized grains decrease. This process leads to the observed enhancement of the levitating grains. Also, these grains are the dominant opacity source for stellar radiation, which peaks at $\sim0.7$~\unit{\micron}, and therefore define the height of the photosphere. As we can see from Fig. \ref{graindist_ratesbehindtrap}, the size of levitating grains increases with height above 2 au due to radiation pressure providing additional vertical support against settling. The grains that make up the enhancement are, therefore, not stationary but slowly being dragged up. The enhancement region can be thought of as a pile-up, a traffic-jam-like effect that is aided by the particulars of the coagulation/fragmentation rates at these heights. As the grains in the overdense region are slowly dragged up, their velocities quickly become large enough to cause fragmentation with the other, smaller grains (again, see Fig. \ref{velocitymags} in appendix \ref{levitatinggrainapp}) - this generates the increased production of 0.1~\unit{\micron} grains above 2.6 au, which are rapidly dragged vertically upwards by the wind. 

We also note that although the over-density of levitating grains is produced at the disc surface, the enhancement is seen at all heights in the disc, all the way down to the mid-plane. This is because, although the grains are levitating with near-zero laminar velocities, the local dust-to-gas ratio where the grains are produced is much larger than at lower heights, so diffusive processes carry dust down the gradient of the dust-to-gas ratio - towards the mid-plane. 

\section{Synthetic observations}\label{obs}

\subsection{Spectral energy distributions}\label{sedsection}

\begin{figure*}
    \centering
    \includegraphics[width=0.99\textwidth]{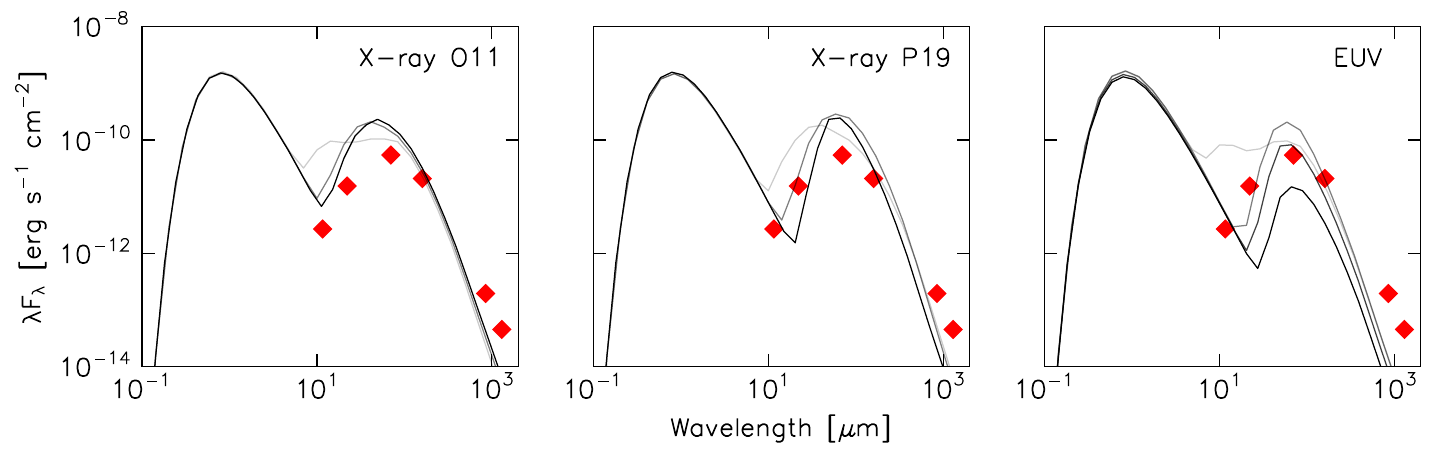}
    \caption{Spectral energy distributions for simulations from each of the different winds models. The line opacity increases with time. The snapshot times after gap-opening for each of these models are; X-ray O11: 0, 0.5 \& 1~Myr, X-ray P19: 0, 0.125 \& 0.25~Myr and EUV: 0, 0.15, 0.25 \& 0.3~Myr. Also plotted in red are observational data for the non-accreting transition disc around IRAS~04125+2902 (data from \citealt{espaillat2015,vdmarel2016}). All systems are taken as being at a distance of 140~pc away from the observer, the distance to the Taurus star-forming region.}
    \label{seds}
\end{figure*}

Fig.~\ref{seds} shows spectral energy distributions (SEDs) calculated for each of the simulations shown in Fig.~\ref{graindistcomp}, one from each wind model. We calculate SEDs using the radiative transfer code \textsc{RADMC3D} \citep{radmc}, using all of the dust species and their associated opacities included in our \textsc{cuDisc} calculations. The SEDs are calculated for an observer inclination of 0$^\circ$ - i.e. a face-on disc. For reference, we include data points from observations of the non-accreting transition disc around IRAS~04125+2902 \citep[see][]{espaillat2015} which typifies the system we are attempting to model. 

We find that the X-ray models maintain a peak mid-to-far IR flux over their entire evolution that is too large to be consistent with observations, whilst the EUV model fits the observed peak after around 0.25~Myr. The X-ray models have all evolved after \HL{the time of }gap-opening for at least 10\% of the primordial disc lifetime, meaning that the longevity of the mid-to-far IR flux is incompatible with the population level statistics of disc observations. As the IR peak is caused by the dust in the trap, this indicates that the X-ray models have too much dust still in the trap; as discussed in section \ref{evores}, this is primarily due to the fact that, whilst each of the wind models have similar mass-loss rates, the initial dust mass in the trap for the X-ray models was larger. The dust mass in the trap for the consistent EUV model is 0.12~\mearth, whilst the X-ray O11 and P19 runs have 0.55 and 0.23~\mearth respectively. These correspond to peak dust-trap surface densities of \num{7.5e-4}, \num{1.7e-2} and \num{1.9e-3}~\unit{\gram\per\cm\squared} respectively. This puts a constraint on the surface density of being $<$\tenpow{-3} to match observed fluxes, which corresponds to $\sim0.1$~\mearth when the trap is at $\sim100$~au. 

We also notice that the far-IR and mm observations show a larger excess, and therefore a shallower slope, than any of the models. This component is due to the dust mass that is in the outer part of the disc (not the trap) and therefore could imply that IRAS~04125+2902 has a larger mass of dust exterior to the trap than in our models. \cite{espaillat2015} fit a disc model with a disc mass that is around an order of magnitude larger than the dust remaining in our discs and an outer radius to the disc of $50-60$~au. This implies that whilst the dust trap must be less optically thick, there should actually be more mass at larger radii; this could potentially be achieved by smaller fragmentation velocities, as these decrease the rate of radially inward dust drift. The difference could also be a function of the dust properties (i.e. opacities and composition), where our chosen properties are not entirely compatible with the conditions in the observed disc. We speculate that models that are able to fit the mm fluxes may actually be better for fitting the near-to-mid IR fluxes, as they imply that less material is able to reach the inner, hotter regions of the disc.

% \begin{figure}
%     \centering
%     \includegraphics[width=0.99\columnwidth]{figures/seds_new2.pdf}
%     \caption{Spectral energy distributions for simulations from each of the different winds models. The snapshot times for each of these models are; X-ray O11: 1~Myr, X-ray P19: 0.25~Myr and EUV: 0.2~Myr. Also plotted in red are observational data for the non-accreting transition disc around IRAS~04125+2902 (data from \citealt{espaillat2015,vdmarel2016}). All systems are taken as being at a distance of 140~pc away from the observer, the distance to the Taurus star-forming region.}
%     \label{seds}
% \end{figure}

\subsection{Scattered light}\label{scatsec}

We also produced scattered light images, again using \textsc{RADMC3D}. These were made by interpolating the vertical dimension to a lower resolution of 80 cells between angles from the mid-plane of $\pm\pi/6$~rad and making the data 3D by copying the ($R,Z$) data into 80 equally-spaced azimuthal cells through $0-2\pi$~rad. These can be seen for the three simulations shown in Figs.~\ref{graindistcomp}~\&~\ref{seds}, at a time of \tenpow{5}~years after gap-opening (the same snapshot as Fig.~\ref{graindistcomp}). 

We see that the EUV disc is already much more optically thin in the disc mid-plane than the other wind models, as the direct stellar emission is clearly visible. We also note that the apparent aspect ratio of the discs differs depending on the wind model; the X-ray O11 model appears flatter than the X-ray P19 model, and the EUV model has a more diffuse appearance with a less discernable aspect ratio. Whilst the gas scale height aspect ratio, $H/R$, is similar for all simulations (0.12 for both X-ray models and 0.14 for the EUV model at 100 au), the photosphere aspect ratios at 100 au are 0.25, 0.3 and 0.22 for the X-ray O11, P19 and EUV models respectively. Since the photosphere is a measure of the height of the small dust, it makes sense that the X-ray O11 model appears flatter than the P19 model given the lower wind mass-loss rate. The EUV model has the lowest photosphere because the dust is extremely diffuse in the outflow due to the low dust mass and large wind velocities; this also explains the diffuse nature of the scattered light image. The EUV model also shows evidence of an X-shaped structure that is associated with the upper surface of the micron-sized dust grains that are entrained in the wind, however the true opening angle of this structure is obfuscated by our $\theta$ domain stopping at angles of $\pi/6$.

% We can compare these structures to recent observations of the edge-on disc Tau 042021, detailed in \cite{duchene2024}, which show an X-shaped feature at 8~\unit{\micron} that they suggest is due to dust entrainment in a wind. In contrast to our results, they do not see this feature at wavelengths of 0.8 and 2~\unit{\micron}, which are instead morphologically similar to the results we find for the X-ray models. An important point to note, however, is that \cite{duchene2024} infer that the dust mass in their observed disc is $>100$~\mearth, making it a young, massive disc as opposed to the transition discs studied in the work. Future work could be to apply our wind and radiation pressure model to young, primordial discs to better compare to these observations and see if we can constrain the wind parameters through such comparisons.
\begin{figure*}
    \centering
    \includegraphics[width=0.99\textwidth]{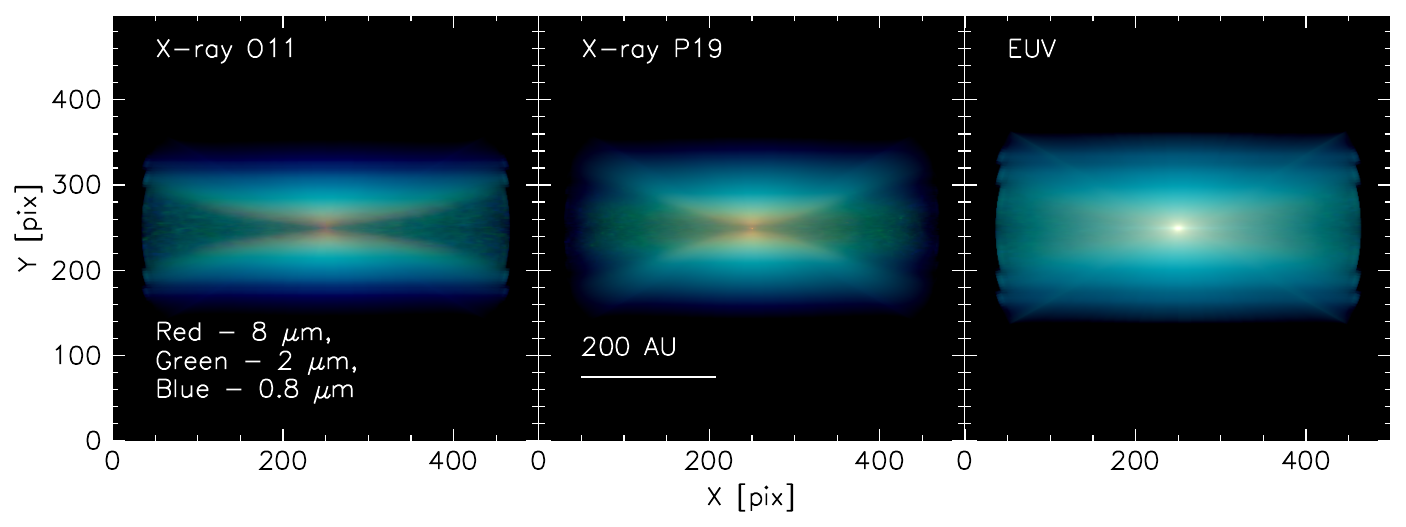}
    \caption{Scattered light images of discs from the three wind models. These false colour images use the intensity of light at 0.8, 2, and 8~\unit{\micron} for the blue, green, and red channels respectively. These particular snapshots are after 100,000~years of evolution.}
    \label{scat1e5}
\end{figure*}

% \begin{figure*}
%     \centering
%     \includegraphics[width=0.99\textwidth]{figures/scatter_comp_end.pdf}
%     \caption{Scattered light images of discs from the three wind models. These false colour images use the intensity of light at 0.8, 2, and 8~\unit{\micron} for the blue, green, and red channels respectively. The snapshot times for each of these models are; X-ray O11: 1~Myr, X-ray P19: 0.25~Myr and EUV: 0.3~Myr.}
%     \label{scatend}
% \end{figure*}

We can compare our results to those found by \cite{franz2022} for dust entrainment in winds. In their work, they used the \cite{picogna2019} wind model, X-ray P19 in our study, so it is of no surprise that the disc morphology we see for this model most closely aligns with their results. However, as our spatial grid does not include angles above the mid-plane that are greater than $\pi/6$ rad, we cannot resolve the cone-shaped features that they observe. However, we can say that the supply of grains to the wind base will be reduced due to the impact of radiation pressure, which would reduce the strength of the cone features. Future work that included this region in our simulations would have to be done to fully understand the impact of radiation pressure on their results, however we also speculate that the opening angle of the cones would increase due to the additional radial force on the grains.

% The images seen at the end of the simulations (Fig.~\ref{scatend}) also show morphological differences. Whilst the X-ray O11 simulation is relatively unchanged, the X-ray P19 simulation shows an X-shaped component related to the dust in the wind, along with a more compact component associated with the dust at lower altitudes. In the EUV model, the small amount of dust that remains is all at high altitudes, with little material at the disc mid-plane, and is all optically thin, meaning that there is little-to-no mid-plane emission. To gain a better understanding of which of these morphologies is truer to reality, we would need to compare to edge-on observations of transition discs, however these are difficult to detect. The lack of these observations makes sense given that transition discs are classified from their SEDs, which can change shape significantly depending on the disc inclination (see \citealt{vdmarel2022}), especially for edge-on discs where the star is mostly obscured. This makes identifying edge-on transition discs a challenging observational endeavour. Future work could be done to identify potential disc candidates that fit this description; observations of such discs could then be used to compare to our models to potentially constrain wind properties.

\section{Discussion}

Given that, in the majority of cases, initial dust masses are too high and mass-loss rates too low for efficient dust removal, the relic disc problem cannot be solved by radiation-pressure-driven outflows, at least not with current photoevaporation models. This implies that a solution may come from the photoevaporation process, and models such as thermal sweeping \citep{owen2012,owen2013,haworth2016} may need to be revisited. The problem could also potentially be solved by processes that reduce the total dust mass at the onset of gap opening, for example sequestration of dust into planetesimals or planets, as the sub-linear relationship seen between dust mass-loss rates and dust surface density in section \ref{dmfluxes} means that lower mass discs can be dispersed on shorter timescales. This is something that could be added to the models presented in the work. Although, as seen in section \ref{sedsection}, lower dust masses may lead to mm fluxes that are too low for observed systems, so some mechanism (e.g. trapping) may meed to be invoked to keep enough dust mass at large radii.

In this work, we have not accounted for the effects of external photoevaporation from the UV background present in star-forming regions (see \cite{winter2022} for a review), nor the difference in gas evolution caused by MHD winds. As opposed to the spreading seen in viscous discs, both of these processes lead to the reduction of the gas disc outer radius over time, MHD winds through direct removal of mass and angular momentum \citep{tabone2022}, and external photoevaporation through direct removal of mass from the outer disc. In general, reducing the gas disc's outer radius would act to drive dust radially inward as the outer pressure slope steepens \citep[see][]{sellek2020}, increasing the flux of dust into the dust trap; this would make it even harder for radiation pressure and wind-driven outflows to remove the optically thick dust from these discs. External photoevaporation can also remove dust from the outer disc via entrainment in the outer wind, especially dust below $10-100$~\unit{\micron} \citep{sellek2020}, which could substantially truncate the dust at radii $>\sim 100$~au in most of our simulations (see fig. \ref{graindistcomp}).

One interesting consequence of the movement of dust from the trap to the outer parts of the disc is that it could potentially explain the measured overabundance of crystalline dust grains at radii exterior to the crystalline radius \citep{vanboekel2005} - the radius where the disc temperature exceeds $\sim800$~K, the temperature at which grains are hot enough to become crystalline. A similar mechanism was invoked by \cite{shu1996} but with grains being deposited at larger radii by MHD-driven winds. In our case, it may be possible that the combination of radiation pressure and either photoevaporative or MHD-driven winds could transport dust grains either from the hot, inner parts of the primordial disc or up to altitudes that allow them to be heated sufficiently by stellar radiation to become crystalline, and then send such grains to large radii in the disc. This problem could be studied by applying our framework to primordial discs and tracking non-crystalline and crystalline grains as multiple dust fluids in the disc.

\HL{Recent hydrodynamic simulations have found that FUV photons can drive photoevaporation from the inner disc in a manner similar to the X-ray models studied here, where X-rays aid the heating but do not constitute the dominant driving radiation field \citep{wang2017,nakatani18,komaki2021}. We did not consider heating by FUV photons for this work because there are currently no works with mass-loss rate fits for photoevaporating discs with inner holes. It would be interesting to study the effect of this in future work, as FUV photons penetrate deeper into the disc atmosphere, moving the wind base closer to the photosphere. We speculate that this could increase dust mass-loss rates as grains that are accelerated by radiation pressure above the photosphere could more easily enter the wind and be lost to the system.}
% We did not consider heating by FUV photons for this work, however we are confident that our results are robust to its inclusion. FUV heating primarily leads to mass-loss in the outer disc \citep{gorti2009}, and the mass-loss rates of dust from the trap depend on the wind profile at and around the inner rim of the disc, meaning that increased mass-loss rates of gas in the outer disc would not affect dust mass-loss. This means that the dust-traps would still remain too optically thick to match observations in mid-to-far IR wavelengths.

One important feature of our findings is that the photosphere height is defined by levitating grains, as discussed in sections \ref{graindist} \& \ref{scatsec}. The photosphere is, therefore, defined by the nature and strength of the wind profile, along with the dust properties. This is important for interpretations of scattered light observations, which trace the spatial distribution of small dust grains. Regardless of the strength of the wind, the photosphere will always lie at larger heights than those that one would calculate for a windless disc. This should be kept in mind when we measure scale heights and degrees of flaring from scattered light observations, as we could misinterpret characteristics of the disc such as the settling efficiency and therefore the degree of vertical turbulent mixing.

\section{Conclusions}

In this work, we have studied how the addition of radiation pressure to photoevaporating transition discs affects dust dynamics, in particular, whether dust can be efficiently removed in a radiation-pressure-driven outflow. The conclusions that we draw from this work are:

\begin{itemize}
    \item In general, sub-micron dust grains are launched along trajectories that point out of the disc system and into the wind, whilst grains larger than a few microns that experience weakened vertical drag from the wind are pushed radially outwards but deeper into the disc interior. In systems with lower dust masses or stronger winds, larger dust grains can be removed in the outflow. 
    \item Radial mass fluxes of the order \tenpow{-7}$-$\tenpow{-6}~\meyr are formed above the photosphere at the radius of the dust-trap. However, the dust trap is also refilled by radial drift of dust grains from the outer disc and by the sweeping up of dust as the inner rim of the gas disc is pushed radially outwards due to photoevaporation. In the majority of cases, the mass-loss rates from the dust-trap are around $5-10~\times$ too small to deplete the trap faster than it can be refilled, leading to the mass in the trap actually increasing as the system evolves. 
    \item Models with EUV winds have lower dust masses at \HL{the time of }gap-opening, which leads to a lower altitude photosphere than the X-ray models. This means larger grains can enter the outflow, and the mass-loss rates are large enough to deplete the dust-trap in timescales that fit observations. \HL{However, atypically high EUV photon fluxes ($\sim$\tenpow{43}~photons~s$^{-1}$) must be invoked.} In addition, whilst the near-to-mid-IR flux constraints are met, the dust masses are too low to match mm fluxes of relic disc observations, implying that either a mass of dust must be kept at large radii (by dust trapping for example), or that another mechanism must be actively removing mass from the dust trap for large mass discs.
    \item Given that most of the parameter space does not lead to efficient dust removal, the solution to the relic disc problem must come from processes that reduce the dust disc mass at \HL{the time of }gap-opening, e.g. planet formation, or from our understanding of photoevaporation.
    \item The shape of the photosphere is governed by the nature of the outflow. Micron-sized grains that are dragged up to the surface layers of the disc define the height of the photosphere, and lead to photosphere flaring that is commensurate with that of passively heated discs, negating any effects of shadowing due to a hot inner rim of the disc. 
    \item An enhancement of `levitating' micron-sized grains is seen due to high relative velocities in the surface layers. The size of these grains is set by the strength of the wind.
    \item Synthetic scattered light images show that the apparent surface of the disc is set by the height of micron-sized grains, i.e. the photosphere, which is controlled by the properties of the outflow as opposed to the gas scale height. This may need to be factored in when interpreting scattered light observations of photoevaporating discs, as the photosphere height is always higher than the gas scale height. This result warrants further study in primordial discs as well as transition discs. 

\end{itemize}

\section*{Acknowledgements}

Richard A. Booth and James E. Owen are supported by Royal Society University Research Fellowships. This work has received funding from the European Research Council (ERC) under the European Union’s Horizon 2020 research and innovation programme (grant agreement no. 853022, Planet Evaporation as a Window into Exoplanetary Origins (PEVAP)) and a Royal Society Enhancement Award. Some of this work was performed using the Cambridge Service for Data Driven Discovery (CSD3), part of which is operated by the University of Cambridge Research Computing on behalf of the Science and Technology Facilities Council (STFC) Distributed Research using Advanced Computing (DiRAC) High Performance Computing (HPC) Facility (www.dirac.ac.uk). The DiRAC component of CSD3 was funded by Department for Business, Energy \& Industrial Strategy (BEIS) capital funding via STFC capital grants ST/P002307/1 and ST/R002452/1 and STFC operations grant ST/R00689X/1. DiRAC is part of the National e-Infrastructure. For the purpose of open access, the authors have applied a Creative Commons Attribution (CC-BY) licence to any Author Accepted Manuscript version arising.

%%%%%%%%%%%%%%%%%%%%%%%%%%%%%%%%%%%%%%%%%%%%%%%%%%
\section*{Data Availability}

The GitHub repository for \textsc{cuDisc} can be found at \url{https://github.com/cuDisc/cuDisc/}. Please feel free to contact Alfie Robinson (a.robinson21@imperial.ac.uk) for any queries relating to the work described in this paper. 



\bibliographystyle{mnras}
\bibliography{references} % if your bibtex file is called example.bib

%\begin{thebibliography}{99}
%\end{thebibliography}

%%%%%%%%%%%%%%%%%%%%%%%%%%%%%%%%%%%%%%%%%%%%%%%%%%

%%%%%%%%%%%%%%%%% APPENDICES %%%%%%%%%%%%%%%%%%%%%

\appendix

\section{Photosphere flaring of different wind models}\label{photapp}

\begin{figure*}
    \centering
    \includegraphics[width=0.99\textwidth]{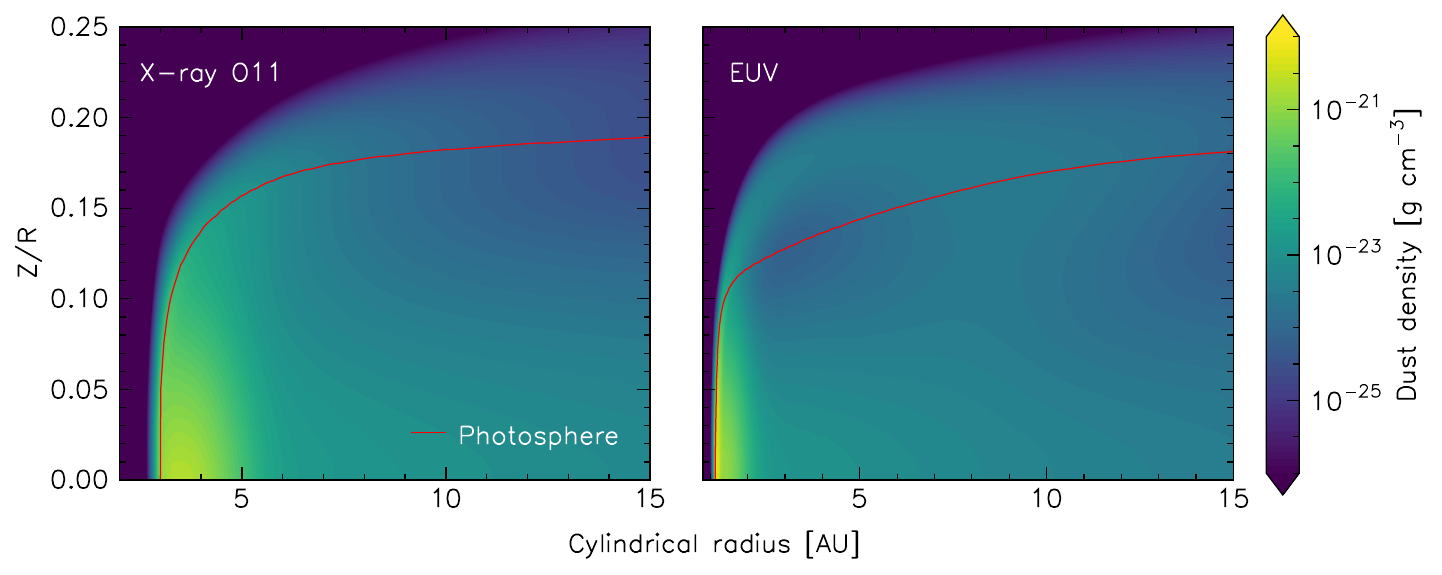}
    \caption{Density profiles of 0.4~\unit{\micron} grains after the outflow has formed for the X-ray O11 and EUV fiducial models. The photosphere (where the optical depth to the peak wavelength of the stellar blackbody is unity) is shown by the red line.}
    \label{photdens}
\end{figure*}

Fig. \ref{photdens} shows the density profiles for sub-micron grains in the fiducial models outlined in section \ref{outflowcharacteristics}. The EUV model exhibits a much more radially confined dust-trap than the X-ray O11 model. This translates to a thin wall of grains that are pulled up at the very inner edge of the disc in the EUV model, compared to a smoother and radially deeper wall in the X-ray O11 model. This difference leads to a deeper photosphere in the first few au for the EUV model, however past $\sim3$~au, the grains trajectories are closer to radial, and the outflow becomes more radially extended, increasing the optical depth.

\section{Levitating grain velocities}\label{levitatinggrainapp}

\begin{figure}
    \centering
    \includegraphics[width=0.99\columnwidth]{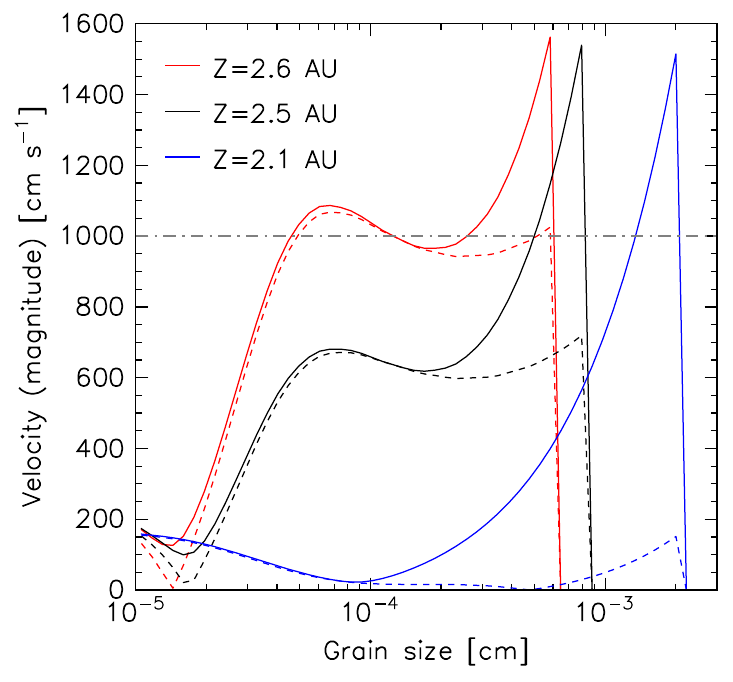}
    \caption{The magnitudes of the grain velocities at three different heights for the simulation shown in fig. \ref{graindist_ratesbehindtrap}. Solid lines show the total velocity (magnitude of both $Z$ and $R$ components) whilst the dashed lines show just the radial velocity. The fragmentation velocity is shown by the grey dash-dotted line.}
    \label{velocitymags}
\end{figure}

Fig. \ref{velocitymags} shows the magnitudes of the velocities of differently-sized grains at three different heights for the simulation shown in Fig. \ref{graindist_ratesbehindtrap}. For heights below the photosphere (2.1~au), we see that relative velocities are small until grains reach sizes of $\sim10$~\unit{\micron}, whilst for heights just above the photosphere (2.5~au) the relative velocities between grains $<1$\unit{\micron} are large, but then become small until the fragmentation velocity is reached for $\sim5$~\unit{\micron} grains. For heights just slightly higher (2.6 au), we see that the velocities of the levitating grains ($\sim0.7$~\unit{\micron}) are now larger than the fragmentation velocity - these are now able to fragment with the low velocity, sub-micron grains. As shown by the dashed lines, the radial velocity due to radiation pressure is key in controlling the relative velocities in this region of the disc.

%%%%%%%%%%%%%%%%%%%%%%%%%%%%%%%%%%%%%%%%%%%%%%%%%%

% Don't change these lines
\bsp	% typesetting comment
\label{lastpage}
\end{document}